\documentclass[10pt,journal,compsoc]{IEEEtran}
\ifCLASSOPTIONcompsoc
  \usepackage[nocompress]{cite}
\else
  \usepackage{cite}
\fi

\ifCLASSINFOpdf
\else
\fi
\usepackage[table]{xcolor}
\usepackage{mathptmx} 

\usepackage{fancyhdr}
\usepackage[normalem]{ulem}
\usepackage[hyphens]{url}
\usepackage[final]{microtype}
\usepackage[keeplastbox]{flushend}
\usepackage{enumitem}
\usepackage[bookmarks=true,breaklinks=true,letterpaper=true,colorlinks,linkcolor=black,citecolor=blue,urlcolor=black]{hyperref}
\usepackage[font=small,labelfont=bf,textfont=bf]{caption}
\usepackage{tabularx, booktabs}
\usepackage{multirow}
\usepackage{tikz}

\usepackage{subcaption}
\usepackage[most]{tcolorbox}
\usepackage{wrapfig}
\usepackage{lipsum}
\newcommand*{\escape}[1]{\texttt{\textbackslash#1}}
\newcommand*{\escapeI}[1]{\texttt{\expandafter\string\csname #1\endcsname}}

\usepackage{algorithm}
\usepackage{algorithmic}
\usepackage{csvsimple}
\usepackage{fixltx2e}

\usepackage{pifont}
\newcommand{\cmark}{\ding{51}}%
\newcommand{\xmark}{\ding{55}}%

\newcommand*\circled[1]{\tikz[baseline=(char.base)]{
            \node[shape=circle,draw,inner sep=0.8pt] (char) {#1};}}

\definecolor{ThisGreen}{HTML}{0bb845}
\newcolumntype{C}[1]{>{\centering\arraybackslash}p{#1}}
\newcolumntype{L}[1]{>{\raggedright\let\newline\\\arraybackslash\hspace{1pt}}m{#1}}
\newcolumntype{R}[1]{>{\raggedleft\let\newline\\\arraybackslash\hspace{1pt}}m{#1}}

\newcommand{\dnum}{34}

\definecolor{commentcolor}{rgb}{0.0,0.0,0.75}

\begin{document}
%
\title{Evaluation of Cache Attacks on Arm Processors and Secure Caches}

\author{Shuwen Deng, Nikolay Matyunin, Wenjie Xiong, Stefan Katzenbeisser, \and Jakub Szefer
\IEEEcompsocitemizethanks{\IEEEcompsocthanksitem S. Deng, W. Xiong and J. Szefer are with the Department
of Electrical Engineering, Yale University, New Haven, CT, 06511.\protect\\
E-mail: \{shuwen.deng,wenjie.xiong,jakub.szefer\}@yale.edu
\IEEEcompsocthanksitem N. Matyunin is with Technical University of Darmstadt, Darmstadt, Hesse, Germany. E-mail: matyunin@seceng.informatik.tu-darmstadt.de
\IEEEcompsocthanksitem S. Katzenbeisser is with University of Passau, Passau, Bayern, Germany. E-mail:  stefan.katzenbeisser@uni-passau.de
}}

\IEEEtitleabstractindextext{%
\begin{abstract}
Timing-based side and covert channels in processor caches continue to be a threat to modern computers. This work shows for the first time a systematic, large-scale analysis of Arm devices and the detailed results of attacks the processors are vulnerable to. Compared to x86, Arm uses different architectures, microarchitectural implementations, cache replacement policies, etc., which affects how attacks can be launched, and how security testing for the vulnerabilities should be done. To evaluate security, this paper presents security benchmarks specifically developed for testing Arm processors and their caches. The benchmarks are themselves evaluated with sensitivity tests, which examine how sensitive the benchmarks are to having a correct configuration in the testing phase. Further, to evaluate a large number of devices, this work leverages a novel approach of using a cloud-based Arm device testbed for architectural and security research on timing channels and runs the benchmarks on 34 different physical devices.  In parallel, there has been much interest in secure caches to defend the various attacks.  Consequently, this paper also investigates secure cache architectures using the proposed benchmarks. Especially, this paper implements and evaluates the secure PL and RF caches, showing the security of PL and RF caches, but also uncovers new weaknesses.
\end{abstract}

}

\maketitle

\IEEEdisplaynontitleabstractindextext

\IEEEpeerreviewmaketitle

\section{Introduction}
\label{sec:introduction}

Over the last two decades, many timing-based attacks 
in processor caches have been exploited to show that it is possible to
extract sensitive information across the logic boundaries established by the software
and even hardware protection mechanisms, e.g.,~\cite{bonneau2006cache, bernstein2005cache, osvik2006cache,
gullasch2011cache, percival2005cache, aciiccmez2006trace}.
Even though a variety of secure processor architectures have been proposed~\cite{szefer2013architectures},
the caches in the proposals are still vulnerable to timing channel attacks.
Further, most recently, Spectre~\cite{Kocher2018spectre} and Meltdown~\cite{Lipp2018meltdown}
attacks have been presented, which attack commercial processors.
Many of their variants depend on cache timing covert channels to extract information.
They exploit speculative
execution to access sensitive data and then make use of 
cache covert channels to actually extract the data.  In most of the attacks,
cache channels are thus critical to actually make the attacks~work.

Despite cache timing channel threats, most of the research has previously focused on x86 processors.
Specifically, there is no previous, systematic evaluation of Arm devices,
despite over $100$ billion Arm processors being 
sold~\cite{link-arm}.

Consequently, this work fills the research gap by analyzing the security of Arm processors
through new security benchmarks developed for testing timing channels in Arm processor caches. 
The benchmarks are built to evaluate
88 types of vulnerabilities previously categorized for processor caches in our conference paper~\cite{deng2020benchmark}.
To gain an understanding of the scope of the vulnerabilities in Arm, this work provides the first, large-scale study of Arm processors,
by testing over \dnum \,different physical devices through three
cloud-based device farms: the Visual Studio App Center~\cite{link-appcenter}, the Amazon AWS 
Device Farm~\cite{link-devicefarm}, and the Firebase Test Lab~\cite{link-firebase}.
For the three cloud-based device farms,
we develop the first {\em cloud-based cache security testing platform}.
We further develop and perform sensitivity tests to evaluate how incorrect
cache configuration information (resulting from misconfiguration or a malicious misinformation)
affects the results of the benchmarks, and which types of tests are most affected by incorrect cache configurations.
As a result, we demonstrate that many of the tests (and attacks),
especially for address-only-based and set-or-address-based vulnerabilities (explained in Section~\ref{sec:sensitivity}), do not require
precise knowledge of the cache configuration.  On the other hand,
this means that attackers can attack the system
even when the cache configuration is unknown -- hiding or intentionally misleading an attacker
about the cache configuration is not a useful defense that one can use.

Compared to our prior conference paper~\cite{deng2020benchmark}, the benchmarking effort in this paper presents new insights
and a number of new solutions we developed to effectively analyze the Arm processors.
Arm uses the {\em big.LITTLE} architecture,
which has heterogeneous caches and CPUs; we are the first to consider this aspect (Section~\ref{subsec:benchmark_heter})
and the first to show the {\em big.LITTLE} architectures provide a larger attack surface
by systematically evaluating different cross-core and cross-CPU 
vulnerabilities in these devices (Section~\ref{subsec:heter_cache}). 
Our work further considers the  pseudo-random replacement policy for caches used by Arm, while our prior
paper~\cite{deng2020benchmark} only considered LRU on x86.  The replacement policy affects the eviction and probing steps used for $48$ out of the 
$88$ types of vulnerabilities and requires new approaches for testing.

Understanding the threats on Arm further requires overcoming a number of challenges.
Cycle-accurate timings are not accessible without root access on Arm, while x86 provides accurate assembly instructions to 
record timing (e.g., {\tt rdtscp}). Our benchmarks closely resemble real attacks by, for example, not assuming root privileges, but 
using code that can get reliable timing in user-level programs.
Our cache timing attack benchmarks use automatically-composed assembly code sequences specialized for Arm.  
This allows for testing different implementations of the assembly for the use in specific attack steps,
and to obtain the final, more accurate vulnerability tests.
We propose the first Arm benchmarks that utilize statistical tests to differentiate distributions of timings to check if 
vulnerabilities can result in attacks, with each benchmark run $30,000$ times to better understand
the timing distributions and minimize potential noise in the measurements.

We also found specific new insights about CPU features affecting security (Section~\ref{subsec:micro_choice}).
For example, we show that the Snoop Control Unit (SCU) in Cortex A53 contains buffers that handle direct cache-to-cache 
transfers; consequently, vulnerabilities related to differentiating cross-core invalidation timing occur much less frequently on 
Cortex A53 than on the other cores.  Meanwhile, the Store Buffer (STB) implemented in Kryo 360 Gold/Silver core pushes the write 
accesses into a buffer, resulting in different timings of accesses to clean and dirty L1 data and resulting in more vulnerabilities.  
These are examples of units that help security, e.g., SCU, and hurt security, e.g., STB.  Only through benchmarking of real devices
can such insights be~discovered.

Given the existing threats due to cache timing attacks, there has already been a number of works on secure caches.
However, none of the cache designs have been systematically evaluated using benchmarks, such as ours.
Consequently, having developed the benchmarks,
we further analyze secure cache designs to understand if they
can enhance security of Arm devices.
This work shows the security of PL~\cite{wang2007new} and RF~\cite{liu2014random} caches, but also uncovers new weaknesses.
Especially, we find a new attack related to eviction-based attacks in the PL cache because it
fails to consider write buffer impacts when locking data in the cache.
Further, we found that the RF cache is secure when setting a large neighborhood window 
(for selecting the randomly fetched cache line). 
A small random-fill neighborhood window, however, may be better for the performance,
but with high probability can leak information about the victim's cache access.

\subsection{Contributions}

In summary, the contributions of this work compared to our prior
conference paper~\cite{deng2020benchmark} are as follows:

\begin{itemize}[noitemsep]
\item Design of the first security benchmark suite and evaluation framework
specifically for Arm, to systematically
explore cache timing-based vulnerabilities 
in Arm devices (considering the {\em big.LITTLE} architecture, pseudo-random cache replacement policy, etc.)
\item Use of a new sensitivity testing approach to evaluate how incorrect
cache configuration information can
affect the benchmarks, and consequently which vulnerability types can still be successful if the cache configuration is incorrect or~unknown.
\item The first large-scale cloud-based test platform
allowing to uncover the security characteristics of a large number of different Arm devices.
\item The first set of cache security benchmarks which can run on the {\em gem5} simulator.
This allows to test microarchitectural features, such as write buffer and MSHR sizes, which cannot be changed on real devices,
and provides an understanding of how they affect the security of the system.
\item Implementation of secure caches in {\em gem5} simulation, and use of the benchmarks to find a new write-based attack on the PL cache 
and problems with the RF cache if the random-fill neighborhood window is not sufficiently large.
\end{itemize}

\subsection{Open-Source Benchmarks}

The Arm benchmarks and the code for the cloud-based framework will be 
released under open-source license and made
available at \url{https://caslab.csl.yale.edu/code/arm-cache-security-benchmarks/}.

\section{Related Work and Background}
\label{sec:related_work}

This section provides background on prior cache timing-based side-channel attacks 
in Arm devices, and gives an introduction to our three-step model used as foundation
for the benchmarks and the evaluation given in this paper.

\subsection{Cache Timing-Based Attacks on Arm}
\label{subsec:previous_arm}

Most of the existing work so far has focused on x86 processors.
For Arm, we are aware of  six 
papers~\cite{lipp2016armageddon,green2017autolock,zhang2016truspy,zhang2016return,haas2021itimed,lee2021hardware} 
that specifically explore security of caches.
Table~\ref{tbl_related_work} lists the related work and compares it to this paper. 
AutoLock~\cite{green2017autolock}  explores how the {\em AutoLock} feature found in some Arm processors
could be used to thwart some cache timing attacks; the paper also shows how attackers can overcome the feature and perform timing attacks.
This work explores previously proposed Evict+Time~\cite{osvik2006cache}, Prime+Probe~\cite{osvik2006cache}, and Evict+Reload~\cite{gruss2015cache} attacks.
ARMageddon~\cite{lipp2016armageddon} focuses on
cross-core cache timing attacks using
Prime+Probe~\cite{osvik2006cache}, Flush+Reload~\cite{yarom2014flush+}, 
Evict+Reload~\cite{gruss2015cache}, and Flush+Flush~\cite{gruss2016flush+} strategies 
on non-rooted Arm-based devices.
TruSpy~\cite{zhang2016truspy}  analyzes timing  cache  side-channel  attacks on Arm TrustZone.
It exploits cache contention between the normal  world  and  the secure  world  to  leak  secret  information from 
TrustZone protected code.
The work only considers the Prime+Probe~\cite{osvik2006cache} attack strategy.
Zhang et al.~\cite{zhang2016return} give
a systematic exploration of vectors for Flush+Reload~\cite{yarom2014flush+} 
attacks on Arm processors and Lee et al.~\cite{lee2021hardware} explore Flush+Reload~\cite{yarom2014flush+} 
attacks on the Armv8 system. 
iTimed~\cite{haas2021itimed} makes use of Prime+Probe~\cite{osvik2006cache}, Flush+Reload~\cite{yarom2014flush+}, 
and Flush+Flush~\cite{gruss2016flush+} to attack Apple A10 Fusion SoC.

While existing works do a good job testing a few vulnerabilities, 
they fail to systematically analyze all possible types of cache timing attacks in Arm processors, as does this work.

\begin{table*}[ht!]
\small
\caption{\small Comparison to related work exploring Arm processors and cache timing attacks.}
  \label{tbl_related_work}
  \begin{center}
  \footnotesize
    \begin{tabular}{| c | c | c | c | c | c |} 
    \hline
     {\bf} & {\bf Num. Vuln. Explored} & {\bf Num. Devices} & {\bf Cloud-Based Framework} & {\bf {\tt gem5} Testing} & {\bf Secure Cache Testing} \\ \hline \hline
     
     AutoLock \cite{green2017autolock} & 3 & 4 & \xmark & \xmark & \xmark \\ \hline 
     
     ARMageddon \cite{lipp2016armageddon} & 4 & 4 & \xmark & \xmark & \xmark \\ \hline 
     
     TruSpy \cite{zhang2016truspy} & 1 & 1 & \xmark & \xmark & \xmark \\ \hline 
     
     Zhang et al. \cite{zhang2016return} & 1 & 5 & \xmark & \xmark & \xmark  \\ \hline 

     Lee et al.~\cite{lee2021hardware}  & 1 & 1 & \xmark & \xmark & \xmark  \\ \hline 

     iTimed~\cite{haas2021itimed}  & 3 & 1 & \xmark & \xmark & \xmark  \\ \hline \hline
     
     {\bf This Work} & 88 & \dnum \, &  \cmark & \cmark & \cmark \\ \hline 
     
     \end{tabular}
  \end{center}
\end{table*}

\subsection{Three-Step Model for Cache Attacks}

Based on the observation that all existing cache timing-based side and covert channel attacks have three steps, 
a three-step model has been proposed previously by the authors~\cite{deng2020benchmark}. 
In the three-step model,
each step represents the state of the cache line after a
memory-related operation is performed. First, there is an
initial step ({\em Step1}) that sets the cache line into a known state.
Second, there is a step ({\em Step2}) that modifies the state of the
cache line. Finally, in the last step ({\em Step3}), based on the
timing, the change in the state of the cache
line is observed.
Among the three steps, one or more steps comprise the victim's access 
to an address that is protected from the attacker (denoted by $V_u$), and timing is observed in {\em Step3}.  
In the model, there are three possible cases for the address of $V_u$: 
(1) $a$, which represents an address known to the attacker, 
(2) $a_{alias}$, which refers to an address that maps to the same cache set as $a$, but is different from $a$,
and (3) Not In Block ($NIB$), which refers to an address that does not map to the same cache set as $a$. 
If a vulnerability is effective, the attacker can infer whether $V_u$ is the same as $a$, $a_{alias}$, or $NIB$  based on the access timing observations.  
The soundness analysis of the three-step model in our prior work~\cite{deng2019analysis} showed that it covers
all possible timing-based attacks in set-associative caches. 
Our recent conference paper~\cite{deng2020benchmark}, upon which this journal paper improves,
presented a benchmark suite based on the three-step model to evaluate 
vulnerabilities in x86 processors -- it did not evaluate Arm processors nor secure cache designs.

\begin{table*}[ht!]
\small
\caption{\small Attack vulnerability types, following~\cite{deng2020benchmark}.}
  \label{tbl_attack_type}
  \begin{center}
  \footnotesize
    \begin{tabular}{| L{2.1cm} | L{13.8cm} |} 
    \hline
     {\bf Attack Type} & {\bf Description}  \\ \hline \hline
     
     $AO$-Type (address-only-based) & In vulnerabilities of this type, the attacker can observe that the timing for victim's access $V_u=a$ is 
     different from the timing for victim's accesses $V_u=a_{alias}$ or $V_u=NIB$, so the attacker can infer if the 
     address of $V_u$ is equal to a known address $a$ or not.
     Vulnerabilities of this type usually differentiate timing between L1 cache hit and DRAM access, 
     which is usually large and distinguishable.  Sample vulnerabilities of this type are
     Flush+Reload (vulnerability benchmarks \#5-\#8 shown in Figure~\ref{fig:device_by_type}).\\ \hline
     
     $SO$-Type (set-only-based) & In vulnerabilities of this type, the attacker can observe that the timing for victim's access
     $V_u=a$ or $a_{alias}$ 
     is different from the timing for victim's access $V_u=NIB$, or the attacker can observe that the timing for victim's access $V_u=a_{alias}$ is different 
     from the timing for victim's accesses $V_u=NIB$ or $V_u=a$. 
     In this case, the attacker can infer the cache set of the address of $V_u$.
     Vulnerabilities of this type usually differentiate timing between L1 cache hit and L2 cache hit, which is usually small.
     Sample vulnerabilities of this type are Evict+Time (vulnerability benchmark \#41 shown in Figure~\ref{fig:device_by_type}).\\ \hline 

	$SA$-Type (set-or-address-based) & In vulnerabilities of this type, the attacker can observe different timing 
	for victim's accesses $V_u=a$, $V_u=a_{alias}$, and $V_u=NIB$. For example, in Prime+Probe (vulnerability \#44), if in 
	{\em Step1}, attacker reads data in address $a$;
	then in {\em Step2}, the victim writes to $V_u$; and then
	in {\em Step3}, the attacker tries to read data in address $a$, data can
	be read from the write buffer (due to the write in the second step if $V_u=a$)
	instead of being read directly from the L1 cache (if $V_u=NIB$ or $V_u=a_{alias}$) and attacker can observe 
	the timing difference of the two cases.\\ \hline\hline

	$I$-Type (internal-based) & Vulnerabilities of this type only involve the victim's behavior in 
	{\em Step2} and {\em Step3} of the three-step model. One example of this attack is the Bernstein's Attack (vulnerabilities \#33-\#36).\\ \hline

	$E$-Type (external-based) & Vulnerabilities of this type are the ones where
	there is at least one access by the attacker in the second or third step,
	e.g., Flush+Reload (vulnerabilities \#5-\#8).\\ \hline

     \end{tabular}
  \end{center}
\end{table*}

\subsection{Cache Vulnerability Types}

We previously identified 88 vulnerability types in caches~\cite{deng2020benchmark}.  To better summarize them, this work categorizes them
into different attack types, as shown in Table~\ref{tbl_attack_type}. 
$AO$-Type (address-only-based), $SO$-Type (set-only-based), and $SA$-Type (set-or-address-based)
categorize the vulnerabilities based on the information that the attacker can gain from the timing 
observation.
Note that our prior work~\cite{deng2020benchmark} 
defined the three types as $A$-Type, $S$-Type, and $SA$-Type, respectively;
we rename the types in this paper to better convey their meanings.
Furthermore, we also categorize them as
$I$-Type (internal-based)  and $E$-Type (external-based) based on whether the 
interference is within the victim process or between the victim process and the attacker process.
These two types of categories are orthogonal to each other. One specific vulnerability
can be both one of $AO$-Type, $SO$-Type, or $SA$-Type, and one of $I$-Type or $E$-Type.
For example, vulnerability \#43 (see Figure~\ref{fig:device_by_type}) belongs to the $E$-$SO$-Type.
Here the E-Type and SO-Type are merged into a combined vulnerability E-SO-Type.

\section{Threat Model and Assumptions}

We assume that there is a victim that has secret data which the attacker tries to
extract through timing of memory-related operations.
The victim performs some secret-dependent memory accesses ($V_u$) and the goal 
for the attacker is to determine a particular memory address (or cache index) accessed by the victim.
The attacker is assumed to have some additional information, e.g., he or she knows the algorithm used by the victim,
to correlate the memory address or index to values of secret data.

In addition to regular reads, writes, and flush operations,
we assume that the attacker can make use of cache coherence protocol to invalidate other core's data, by triggering
read or write operations on the {\em remote} core as one of the steps of the that attack.

A negative result of a benchmark means there is likely
no such timing channel in the cache or the channel is too noisy to be observable. 
Meanwhile, a positive result may be due to structures other than cache, 
such as prefetchers, Miss Status Handling Registers (MSHRs), 
load and store buffers between processor and caches, or line fill buffers between cache levels.
Our benchmarks focuses on L1 data caches, but we consider that timing results could
be due to all the different structures.
Detailed benchmarks for these structures or other levels of caches are left for future work.

\section{Arm Security Benchmarks}
\label{sec:benchmark}

In this section, we present the first set of benchmarks
which is used to evaluate L1 cache timing-based vulnerabilities of Arm processors.  
To implement the security benchmarks on Arm,
as listed below, we developed solutions to key challenges accordingly.

\subsection{Heterogeneous CPU Architectures}
\label{subsec:benchmark_heter}

Arm processors implement the {\em big.LITTLE} architecture 
with {\em big} and {\em little} processor cores having different cache sizes.
This presents a new challenge,
as the architecture is fundamentally different from multi-core systems where all cores
have identical cache sizes and configurations.
This was not considered in our previous work~\cite{deng2020benchmark} which only dealt with x86,
nor in previous studies~\cite{lipp2016armageddon,green2017autolock,zhang2016truspy,zhang2016return}
which only tested attacks on one core type.

The {\em local} core is the one wherein is located the target cache line
that the attacker wants to learn.  Meanwhile, the {\em remote} core is a different core where the target cache line is not located,
but which could affect the {\em local} core and its caches, e.g., via cache coherence protocol.
Thus, both cross-core and cross-CPU vulnerabilities are considered in our work by testing the victim and attacker
operations on different combinations of {\em local} and {\em remote} cores.
Especially, with different {\em big} and {\em little} processor cores, 
a {\em local} or {\em remote} core can be either of {\em big} or {\em little} core type, resulting in four combinations.

Because we consider different core types, unlike prior work,
and caches are not even between the {\em big} and {\em little} cores,
we define how to correctly specify the cache configurations for the benchmarks when running the tests:

\begin{itemize}[noitemsep]
  \item If the first two steps of the three-step model describing a particular vulnerability both occur in the remote core, use the remote core's cache configuration.
  \item In all other cases, use the local core's cache configuration.
\end{itemize}

In the three-step model, when testing for vulnerabilities, main interference (leading to potential timing differences)
occurs within the first two steps, while the final, third step is used for the timing observation
used to determine if there is possible attack or not.
Therefore, 
the above method of choosing the cache configuration focuses on 
where the main interference is occurring in the three steps.

\subsection{Random Replacement Policy in Arm}

Modern Arm cores use the random replacement policy in the L1 cache~\cite{lipp2016armageddon}. 
This policy is significantly different from the Least Recently Used (LRU) replacement policy, and poses fundamental challenges for 
eviction and probing steps in $48$ out of $88$ vulnerability types.

In particular, this makes the set-only-based vulnerabilities ($SO$-Type) harder to implement.
The reason is that occupying a cache set in caches using a random replacement policy is not as easy as in caches 
using LRU or similar policies, where accessing a certain 
number of ways (denoted as {\tt cache\_associativity\_num})
of cache lines in a cache set is able to evict all data in the set.
In caches using the random replacement policy, the
{\em cache set thrashing problem}~\cite{tromer2010efficient}, referring to self-evictions within the eviction set,
which affects accessing all the ways of the cache set in eviction-based vulnerabilities.
To avoid this problem, we use a smaller
set size to avoid set thrashing in our benchmarks. 
We set the eviction set size to  {\tt cache\_associativity\_num-1} and then 
repeat each step's memory operations 10 times.
Using this technique, we are able to reduce  
set thrashing significantly given the random replacement policy.
However, in this case, exactly one way
will not be occupied after the repeated memory operations.
This will cause victim's access in one out of {\tt cache\_associativity\_num} ways to be not detectable, but this is acceptable
as vulnerabilities can still be detected as we show in our evaluation.

\begin{figure}[t]
\centerline{\includegraphics[width=7.5cm]{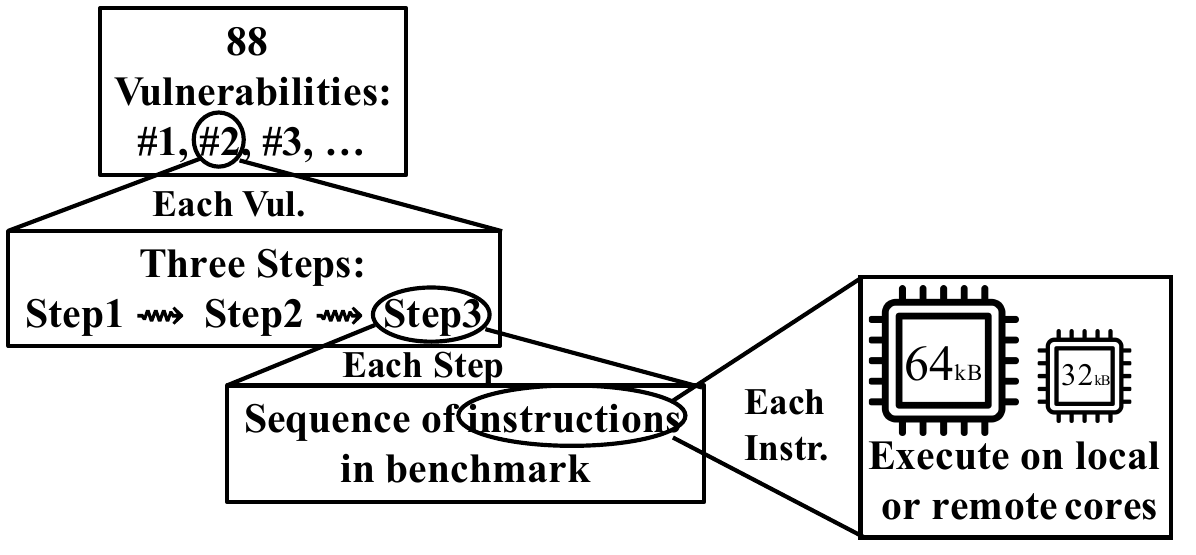}}
\caption{\small Relationship of the 88 vulnerabilities, each of which is described using three steps
from the three-step model. The steps are further translated into sets of assembly instructions for the benchmarks, and the
code can be run on either {\em big} or {\em small} cores in the tested systems.}
\label{fig_from_three_step_to_benchmarks}
\end{figure}

\subsection{Measuring and Differentiating Timing}

For benchmarking Arm cache timing-based vulnerabilities, this work is the first to utilize 
statistical tests -- Welch's t-test~\cite{welch1947generalization} -- 
to differentiate distributions of timings to check if vulnerabilities can result in attacks.
The {\em pvalue} is the threshold used to judge the effectiveness of the vulnerabilities.
Based on our evaluation, we select $0.00049$ for the {\em pvalue} in our tests, improved from our previous work on x86~\cite{deng2020benchmark},
and use this to determine if different timing distributions are distinguishable.
We chose Welch's t-test since it is generally used in attack evaluations~\cite{bhattacharyya2019smotherspectre,bhattacharya2020exploring,crosby2009opportunities}.
There is also Kolmogorov-Smirnov's two-sample test~\cite{smirnov1939estimation}
that can be used to differentiate distributions.

The statistical tests are used to differentiate timings of memory related operations.
However, cycle-accurate timings are not accessible without root access on Arm, 
while x86 provides accurate assembly instructions to record timing (e.g., {\tt rdtsc}). 
Consequently, we developed code that can get reliable timing measurements 
in user-level applications using the {\tt clock\_gettime()} system call.
We experimented with other different performance counters and thread timers,
but these proved not to be applicable or accurate enough for our benchmarks.

When performing timing measurements, in our experience, Arm devices further exhibit a lot of system noise 
when running the tests on real devices in the cloud-based device farms,
possibly due to OS activity, or other background services.
Therefore, we set the benchmarks to run more than $30,000$ repetitions for each benchmark for each device
to average out the noise.
Further, when running each step operated by either the victim or the attacker, we isolate the core 
to avoid influence of other application processes from user-level applications.

\begin{algorithm}[!t]
\small
\caption{\small Read/Write Access Code Sequence}
\label{alg:A}
\begin{algorithmic}[1]

\STATE asm \_\_volatile\_\_ (
\STATE      ``DSB SY            \escape{n}'' 
\STATE      ``ISB               \escape{n}''  
\STATE      ``LDR/STR \%0, [\%1]     \escape{n}'' 
\STATE      ``DSB SY            \escape{n}''  
\STATE      ``ISB               \escape{n}''  
\STATE      : ``=r'' (destination)
\STATE      : ``r'' (array[i]));

\end{algorithmic}
\end{algorithm}

\begin{algorithm}[!t]
\small
\caption{\small Flush Code Sequence}
\label{alg:B}
\begin{algorithmic}[1]

\STATE asm \_\_volatile\_\_ (
\STATE      ``DSB ISH            \escape{n}'' 
\STATE      ``ISB               \escape{n}''  
\STATE      ``DC CIVAC, \%0     \escape{n}'' 
\STATE      ``DSB ISH            \escape{n}''  
\STATE      ``ISB               \escape{n}''  
\STATE      : : ``r'' (array[i]));

\end{algorithmic}
\end{algorithm}

\subsection{Summary of Benchmark Structure}

Following the above features, we developed benchmarks for all 88 vulnerabilities.
As shown in Figure~\ref{fig_from_three_step_to_benchmarks},
there are three steps for each vulnerability, and
each step is realized by a sequence of instructions.
The instruction sequences from each step can execute on local or remote cores.
When performing the steps, there are two possible cases for the victim's or attacker's memory related operation:
read or write access for a memory access operation;
and flush or write in the remote core for an invalidation-related operation.
Thus, for each vulnerability, there are in total of $2^3=8$ types considering 
different cases of each step's operation.
Further, if a vulnerability being tested has both the victim and the attacker running on one core, these two parties can run either time-slicing or multi-threading. 
Consequently, the 8 cases are doubled to account for both time-slicing and multi-threading execution.
Thus, for each vulnerability being tested, there are correspondingly 8-16 cases depending on the specific vulnerability.
Each vulnerability is realized as a single benchmark program.
In total there are 1094 benchmarks for all 88 types of vulnerabilities.

The 1094 benchmarks are automatically generated.   The basic code sequences, e.g., Alg.~\ref{alg:A} and~\ref{alg:B}, are
composed into programs, with one program for each benchmark.  Additional instructions are used in the benchmarks to pin execution
of the code to different processor cores when testing different configurations.  
The resulting 1094 programs are compiled and executed on the devices under test
as detailed in the next section.

\section{Cloud-Based Framework}
\label{sec:framework}

In this section, we report on the first cloud-based platform for testing cache channels on Arm devices. 
Our prior work only considered x86~\cite{deng2020benchmark} with several processors manually set to test, and
work by others only 
manually tested only few Arm devices~\cite{lipp2016armageddon,green2017autolock,zhang2016truspy,zhang2016return}.

\subsection{Android Device Testbeds}

We build our evaluation framework using testing platforms for mobile devices, namely
the Visual Studio App Center~\cite{link-appcenter}, the Amazon AWS 
Device Farm~\cite{link-devicefarm}, and the Firebase Test Lab~\cite{link-firebase}.
We developed a framework which allows us to run custom binary benchmarks and retrieve the 
results in an automated manner. 

\begin{figure}[t]
\centerline{\includegraphics[width=6.5cm]{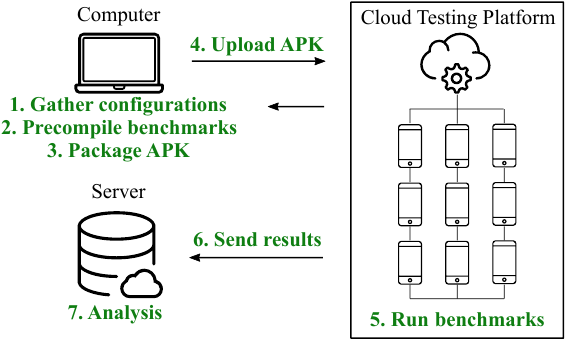}}
\caption{\small Overview of the evaluation framework using the cloud-based testing platforms for Android mobile devices. }
\label{fig:scenario}
\end{figure}

In these cloud deployments, it is not possible to execute benchmark files through
a remote shell and download the results. Instead, the entire functionality 
must be implemented as a user-level native Android application.
Consequently,
the benchmark executables are inserted into
the application package (APK) of a custom Android application we developed.
Figure~\ref{fig:scenario} illustrates the resulting test setup, which will be open-sourced.

\subsection{Extracting Cache Configurations}

To build the benchmark, cache and CPU configuration information are needed.
The configuration can be automatically identified 
by reading the corresponding system information located at \textit{/sys/devices/system/cpu/cpu$x$/} (where $x$ stands for the CPU core number)
on each tested device.
However, depending on the SELinux policies applied by the vendor and Android version,
access to these files is restricted on some devices~\cite{link-selinux}.
For these device models, we manually identify and verify their cache configurations from public resources.
Finally, we store both automatically- and manually-extracted cache configuration parameters in a single database,
and include this database into the APK, so that it can be used when running the benchmarks.

\subsection{Packaging Security Benchmarks}

Starting from Android 9, the operating system does not allow to execute files from an arbitrary
writeable location on the filesystem~\cite{link-exec-restrictions}.
Instead, only native library dependencies within an Android application can be executed.
Consequently, we pre-compile and place the benchmark files in the resource subfolder of
the APK package which contains native libraries (\textit{src/main/resources/lib/arm64-v8a}), as the OS grants read-and-execute permissions
for all binary files in this subfolder.

\subsection{Running Benchmarks}

We give an overview of our evaluation framework in Figure~\ref{fig:scenario}.
Once the cache configuration is extracted (step 1), 
the corresponding benchmarks are  precompiled (step 2) and packaged  (step 3), we upload the application package to the cloud testing platforms (step 4).
The implemented application does not require
any user interaction. Instead, it contains an instrumented unit test which automates the execution of benchmarks.
The tests can be run simultaneously on multiple devices  (step 5).
The process of uploading and running the application is automated using the APIs provided by the cloud platform provider. 

On each device, the application first identifies the device model by accessing the \textit{Build.MODEL} property. 
This information is used to look up the corresponding cache configuration parameters in the database.
Afterwards, the application executes the precompiled benchmarks one by one, using the corresponding parameters.
In order to automatically retrieve the results of benchmarks from multiple devices, we implement an HTTP server which can receive
POST requests from Android applications. Each request contains the results in textual or binary format. As the execution time of the whole
set of benchmarks on a device can take several hours, the application periodically sends the intermediate results to the server. In this way,
we can precisely monitor the current state of the execution on each device.
Finally, the results are collected from the server  (step 6) for further analysis  (step 7).

\section{Evaluation}
\label{sec:evaluation}

\begin{figure*}[t!]
\centering
\includegraphics[width=\textwidth]{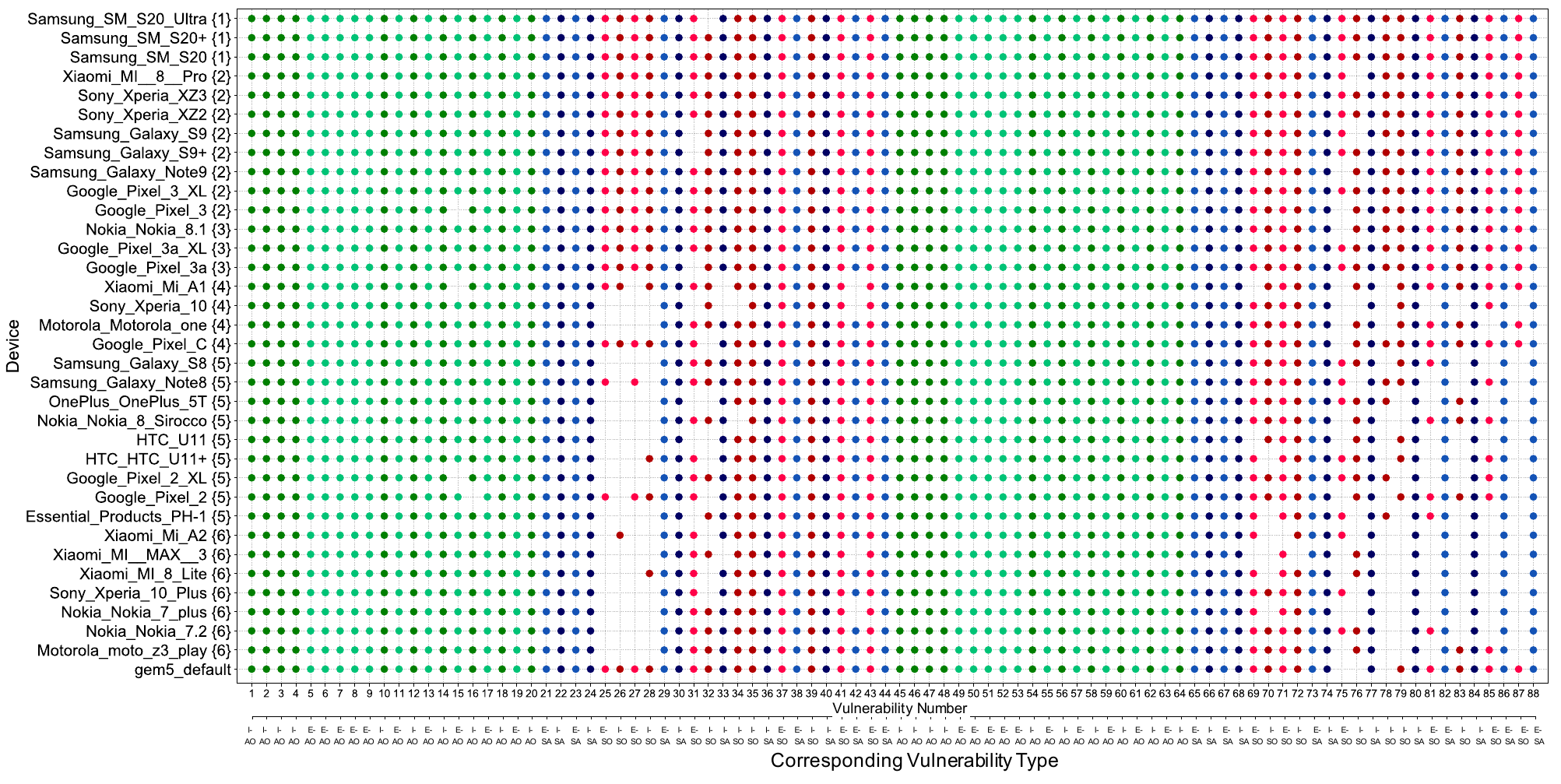} 
\caption{Evaluation of the 88 types of vulnerabilities on different Arm devices.
A solid dot means the corresponding processor is found to be vulnerable to the vulnerability type.
The ``$I$-$SO$'' (colored by dark red) and ``$E$-$SO$'' (colored by light red) 
are internal-interference set-only-based and 
external-interference set-only-based vulnerabilities, respectively.
The ``$I$-$AO$'' (colored by dark red) and ``$E$-$AO$'' (colored by light red) 
are internal-interference address-only-based and 
external-interference address-only-based vulnerabilities, respectively.
The ``$I$-$SA$'' (colored by dark red) and ``$E$-$SA$'' (colored by light red) 
are internal-interference set-or-address-based and 
external-interference set-or-address-based vulnerabilities, respectively. 
The devices are grouped according to their core types. Each device's core is labeled by a number shown after the device name, with corresponding cores shown in Table~\ref{tbl:setup}. 
The order is from the most vulnerable core to least vulnerable among the cores. 
The last line shows {\tt gem5} testing results of default {\tt gem5}, to show that {\tt gem5} simulation gives similar results to real devices.
}
\label{fig:device_by_type}
\end{figure*}

\begin{table}[t!]
\centering{
{\fontsize{7.5}{9}\selectfont
\csvreader[tabular=|C{1.336cm}|C{1.07cm}|C{1.61cm}|C{2.25cm}|C{0.415cm}|,
  table head=\hline Core Name	& Core Freq.& 	L1 Cache Config.&	SoC Name&	Vul. Num.\\\hline\hline,
  column count=5,
  late after line=\\\hline]%
  {figures/core_info.csv}{}%
{\csvlinetotablerow}%
}
\vspace{0.2cm}
\caption{\small CPUs and SoC types found in the evaluated devices.
The {\em Core Name} (with corresponding number used in Figure~\ref{fig:device_by_type}), {\em Core Freq.}, and {\em L1 Cache Config.} columns
show the processor core names, their frequency ranges, and typical cache configurations.
The {\em Vul. Num.} column shows the average number (out of 
88) of vulnerabilities that show up during tests;
smaller value is better.
}
\label{tbl:setup}
}
\end{table}

We tested a total of \dnum \, different devices.
The corresponding processor core types are shown in Table~\ref{tbl:setup} -- note 
that some devices use the same processor or SoC 
configuration so there are less than \dnum \,processors in Table~\ref{tbl:setup}.
The results of the tests are shown in Figure~\ref{fig:device_by_type},
which shows the vulnerabilities that can possibly be exploited on the device, based on sufficient
timing differences in the memory operations corresponding to each three-step attack.
Figure~\ref{fig:device_by_type} consists of $88$ columns, each corresponding to one 
of the three-step vulnerabilities.  
The vulnerabilities are colored based on the different types.

In addition to smartphones, we further tested other Arm cores, leveraging Amazon
EC2~\cite{amazon} with an X-Gene 2 core and Chameleon cloud~\cite{keahey2020lessons} 
with a Neoverse core to test Arm processors on
servers.
Arm server chip results generally have similar patterns as the
mobile devices. 
Therefore, in this work, we show only results for the mobile devices from the cloud-based testbeds.

\subsection{Microarchitectures' Impacts on the Vulnerabilities}
\label{subsec:micro_choice}

Below we list some of the observations gained from our evaluation.
Only through the extensive benchmarking of caches on a large set of devices, can such insights be discovered.

\subsubsection{Store Buffer}

The STB (STore Buffer) 
is used during write accesses to hold store operations.
This structure makes 
clean and dirty L1 data access timing easier to be distinguished.
For example, $I-SA$-Type vulnerability \#33 differentiates 
timing between reads of dirty L1 data and reads of clean L1 data, or 
between writes of dirty L1 data and writes of clean L1 data, which is a typical vulnerability that 
allows STB to make it more effective.
From the evaluation results, Kryo 360 Gold/Silver cores are more susceptible to vulnerabilities such as \#33,
compared to Cortex A53 core, which confirms the fact that
the STB is presented in Kryo 360 Gold/Silver cores but not in Cortex A53 core, based on reference manuals.

\subsubsection{Snoop Control Unit}

The Snoop Control Unit (SCU) contains buffers that can handle direct cache-to-cache transfers between cores without having
to read or write any data to the lower cache
by maintaining a set of duplicate tags that permit each coherent data request to be
checked against the contents of the other caches in the cluster.
With the SCU, when comparing the timing between 
remote writes to invalidate local L1 data and remote writes to invalidate local L2 data,
the SCU will accelerate the coherence operations. This  makes
the different cache coherence influence non-differentiable in timing
on the cores that have the SCU.

For example, $I-SO$-Type vulnerabilities \#78-\#79
mainly use timing differences between
flushing of L1 data and flushing of L2 data, or between
remote writes to invalidate local L1 data and remote writes to invalidate local L2 data. 
From the evaluation results, 
vulnerabilities \#78-\#79
occur much less frequently
on Kryo 280 Gold/Silver cores and Cortex A53 cores compared to Kryo 360 and Kryo 385 Gold/Silver cores. 
This supports the observation that
the Kryo 280 Gold/Silver cores and Cortex A53 cores
have a Snoop Control Unit (SCU), which helps prevent these types of vulnerabilities, while Kryo 360 and Kryo 385 Gold/Silver cores do not have it.

\begin{figure*}[t!]
\centering
\includegraphics[width=\textwidth]{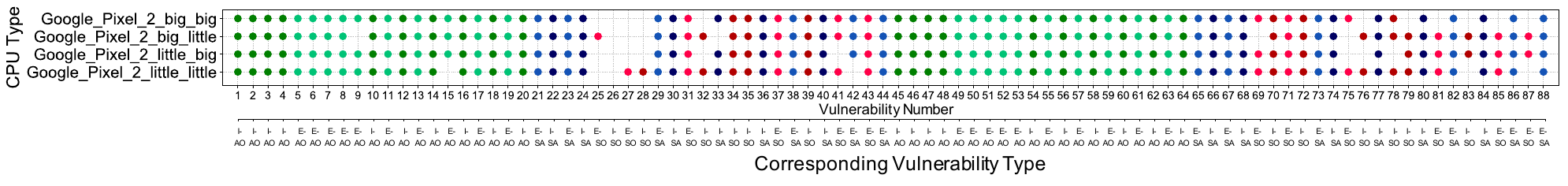} 
\caption{Evaluation of the 88 types of vulnerabilities on different cores of Google Pixel 2.
``big\_big'' means running both local and remote core on big cores, ``big\_little'' 
means running local core on the big core, remote core on the little core. 
Same naming is applied to ``little\_big'' and ``little\_little''.
Dot coloring is the same as in Figure~\ref{fig:device_by_type}.}
\label{fig:core_test}
\end{figure*}

\subsubsection{Transient Memory Region}

Transient Memory Region allows for setting a memory region as transient.
Data from this region, when brought into L1 cache, will be marked as transient.
As result, during eviction, if this cache line is clean, it will be marked as invalid instead of being allocated in the L2 cache.

Although this may help avoid polluting the cache with unnecessary data, 
internal and external $SO$-Type vulnerabilities \#33-\#44 that we are able to differentiate between L1 
and L2 cache hits can now differentiate between an L1 cache hit and a data access from DRAM.
This makes this type of vulnerability more effective on 
cores that support this feature, which are Kryo 360/385 Gold/Silver cores, compared to other cores, such as Cortex A53.

\begin{figure}
\centering
\begin{subfigure}[b]{0.47\textwidth}
   \includegraphics[width=\linewidth]{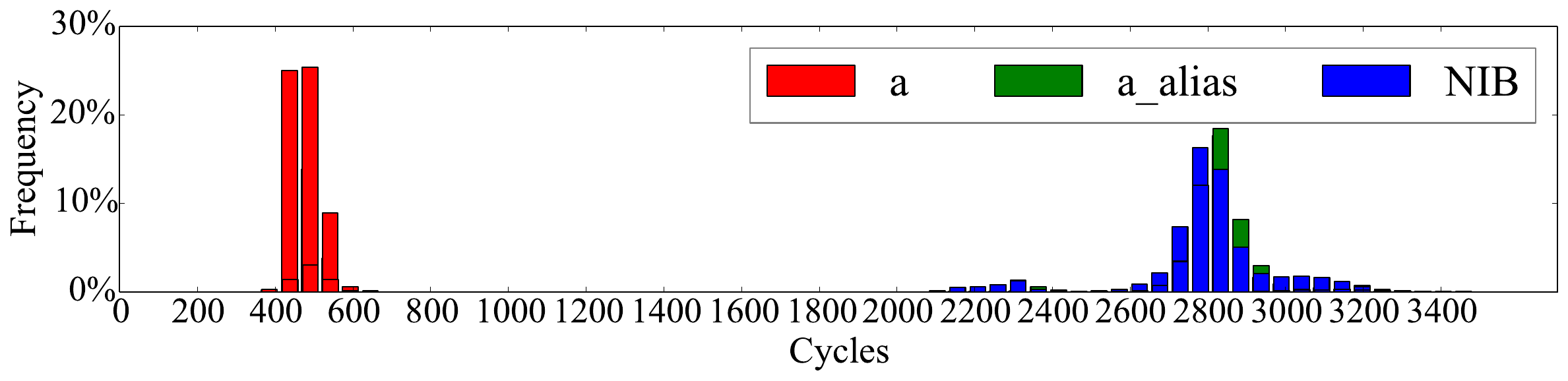}
   \caption{Sample histogram of $AO$-Type vulnerability}
   \label{fig:Ng1} 
\end{subfigure}

\begin{subfigure}[b]{0.47\textwidth}
   \includegraphics[width=\linewidth]{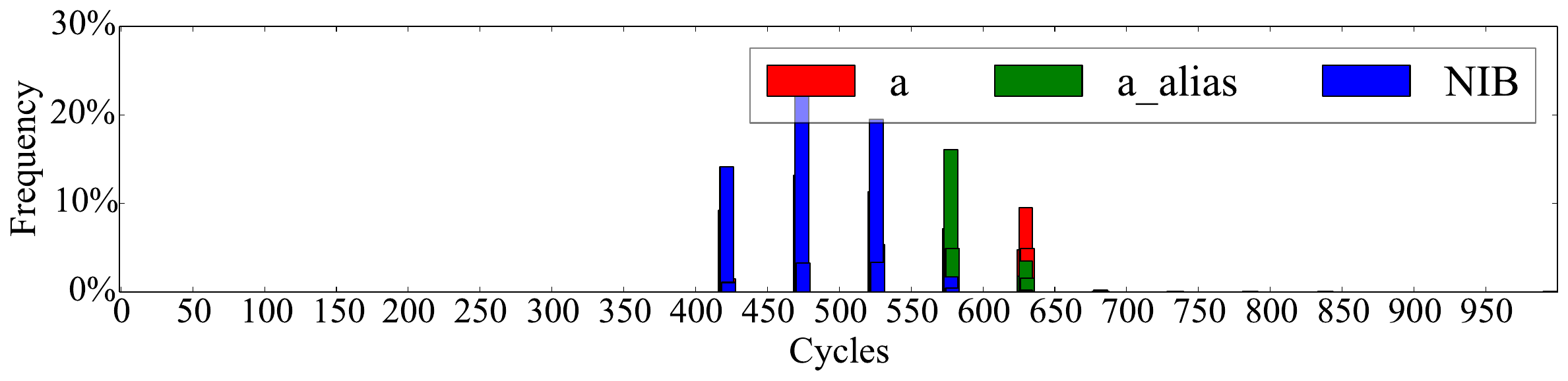}
   \caption{Sample histogram of $SO$-Type vulnerability}
   \label{fig:Ng2}
\end{subfigure}

\caption{Samples of different types of vulnerabilities'  timing histograms for different candidate values for $V_u$.}
\label{fig:hist} 
\end{figure}

\subsection{Heterogeneous Caches' Impact on Vulnerabilities}
\label{subsec:heter_cache}

We also evaluated how Arm's {\em big.LITTLE} architecture impacts the attacks, 
where we set {\em local} and {\em remote} core to be either {\em big} or {\em little} processor core. 
In Figure~\ref{fig:core_test}, we present evaluation results for one example device, Google Pixel 2.
A similar pattern was observed for all other tested devices.

$SO$-Type and $SA$-Type vulnerabilities which differentiate 
L1 and L2 cache hit timings (\#41-\#44) are mostly vulnerable to the case 
when the {\em local} core uses the {\em big} core.
This is mainly because the bigger cache (e.g., 64K 16-way vs. 32K 4-way) of the {\em big} core
results in larger timing differences for the vulnerabilities that require priming each cache set, 
reducing the proportion of system noise at the same time.
$SO$-Type and $SA$-Type vulnerabilities which differentiate 
writing to remote dirty L1 and L2 cache data (\#73-\#76) are successful
when {\em local}  and {\em remote} core both use the {\em little} core.
Dirty data are usually not stored in the cache line but stored in other locations
such as write buffer.
Write buffer is possibly processed in an out-of-order way. 
Therefore, fewer number of writes due to fewer number of ways in {\em little} core 
are more likely to have relatively differentiable timing.
$SO$-Type and $SA$-Type vulnerabilities which differentiate writing remote 
L1 and remote L2 cache data (\#77-\#88) are mostly successful
when {\em local} and {\em remote} cores use different core types ({\em big} or {\em little}).
This is due to the fact that {\em big} and {\em little} cores are often in different quad-core clusters in the SoC, where 
coherence time across quad-core cluster results in higher timing differences when accessing data located in the remote cluster.

\subsection{Core Frequency's Impact on Vulnerabilities}

High clock frequency tends to make long memory operations more differentiable,
and will make timing attacks easier to exploit the difference.
From the evaluation results, we found that
devices with higher clock frequency will likely have more effective timing-channel vulnerabilities. 

This is especially visible in $SO$-Type vulnerabilities, most of which differentiate between 
L1 and L2 cache hits, which have a relatively small cycle difference, e.g., less than 10 cycles.
However, if the core's frequency increases, the timing difference is 
also increased, which makes cycle distributions more differentiable and an attack possibly easier to execute.

\begin{figure*}
\centering
\begin{subfigure}[b]{\textwidth}
   \includegraphics[width=\textwidth]{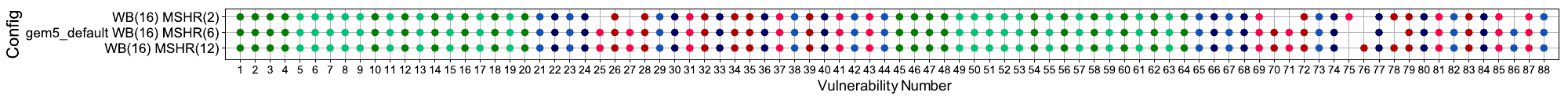} 
   \caption{Benchmark {\tt gem5} simulation results for different MSHR sizes.}
\end{subfigure}

\begin{subfigure}[b]{\textwidth}
   \includegraphics[width=\textwidth]{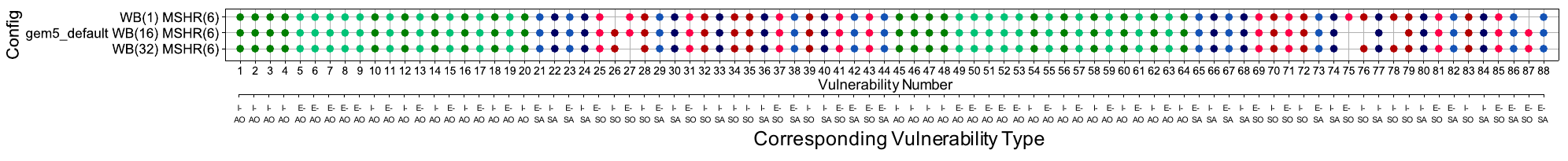} 
   \caption{Benchmark {\tt gem5} simulation results for different write buffer sizes.}
\end{subfigure}

\caption{\small Evaluation of 88 types of vulnerabilities on different number of write buffer (WB) and MSHR sizes.
A solid dot means the corresponding processor is found to be vulnerable to the vulnerability type. 
The ``$SO$'' (colored red) and ``$AO$'' (colored green) are set-only-based and address-only-based vulnerabilities, respectively.
``$SA$'' (colored blue) are the ones that are set-or-address-based.
The ``$E$'' (colored in lighter color) and ``$I$'' (colored in darker color) are internal- and 
external-interference vulnerabilities, respectively.  }\label{fig:gem5_all}
\end{figure*}

\subsection{Influence of Write Buffer and MSHR Sizes}
\label{subsec:gem5_wb_mshr}

We design our benchmarks so they can also be used in simulation.
We use the Arm {\em big.LITTLE} configuration
to run the benchmarks in Full System (FS) mode or Syscall Emulation (SE) mode on {\tt gem5}.
The simulator is configured to use the Exynos~\cite{burgess2016samsung} configuration to model real Android
devices and uses the O3CPU model with a 5-stage pipeline. 
The last line of Figure~\ref{fig:device_by_type} shows the benchmark results when using the default configuration on the {\tt gem5} simulator.
Overall, we find that baseline {\tt gem5} results have good correspondence with real CPUs in terms of the cache timing vulnerabilities.

Next, we evaluate different configurations of the Miss Status Holding Register (MSHR) and the write buffer (WB),
both tested on {\tt gem5}. 
Results are shown in Figure~\ref{fig:gem5_all}:
A larger MSHR size  leads to more 
vulnerabilities to be observed.
MSHR is a hardware structure for tracking outstanding misses.
Larger MSHR sizes lead to more outstanding misses that can be handled, which may stabilize
the memory access timings and give more consistent results.

Changing the size of WB does not have an explicit influence on the vulnerability results. 
WB stores the write request, which frees the cache to service read requests while the write is taking place. 
It is especially useful for very slow main memory, 
where subsequent reads are able to proceed without waiting.
We use the ``SimpleMemory'' option of {\tt gem5}, which is relatively simple compared 
with the implementation of real devices and may not have the same slow memory timing in this case. 
As the result shows, bigger WB may improve performance and can be added without degrading security, while bigger
MSHR may improve
performance but at some cost to security.

\subsection{Patterns in Vulnerability Types}
\label{subsec:patterns}

It can clearly be observed from the colored dots in Figure~\ref{fig:device_by_type} that
$AO$-Type vulnerabilities are observable in almost all devices and in the simulation, 
because these types of vulnerabilities, e.g., differentiate L1 cache hits and DRAM hits, which have large timing differences. 
Such timing distribution results can be observed in Figure~\ref{fig:Ng1}.
$SA$-Type vulnerabilities  also occur relatively often, but are much more unstable compared with $AO$-Type vulnerabilities,
which shows that different devices have large but quite variable timing differences among different memory operations,
e.g., between clean abd dirty L1 data invalidation 
or between local access of remote clean and dirty L1 data.
$SO$-Type vulnerabilities are least effective.
This is because the timing differences between the observations such as
L1  and L2 cache hits are so small that they are sometimes indistinguishable due to 
system noise.
For example, 
timing distribution evaluation result shown in Figure~\ref{fig:Ng2} have small timing difference.

$I$-Type and $E$-Type vulnerabilities do not show explicit evaluation differences.
In this case, another take-away message is that protecting only the external-interference 
vulnerabilities is not enough at all.
Internal-interference vulnerabilities can be as effective as the external-interference 
vulnerabilities for attacks.

\subsection{Estimating the Real Attack Difficulty}

To estimate the real attack difficulty, we can leverage the distance and likelihood (using p-value) of different timing measurement distributions.
As is shown in Figure~\ref{fig:hist} in Section~\ref{subsec:patterns}, $AO$-Type or $SA$-Type vulnerabilities are easier to exploit since 
they depend on timing differences of L1 
cache hits vs. DRAM accesses; meanwhile $SO$-Type vulnerabilities are more difficult to exploit, since they depend on the timing differences
between L1 and L2 cache hits, which are much smaller compared to the former.

Further, our benchmarks show the overall attack surface.
If a motivated attacker only needs to use one attack to derive sensitive information, he or she will likely start
with $AO$-Type or $SA$-Type vulnerabilities.
However, the bigger the attack surface is, the more options he or she has, and if there are defenses for $AO$-Type or $SA$-Type types of vulnerabilities,
attackers could still leverage $SO$-Type vulnerabilities.
The goal of this work is to show the whole attack surface on Arm devices,
including vulnerabilities and attack types that are not previously presented in the literature.
Which attack could be used in practice depends on the attacker's motivation and resources, but thanks to this work,
the overall attack surface is better understood.

\subsection{Results Compared with Other Work}

For our benchmark results shown in Figure~\ref{fig:device_by_type}, 
strategies exploited by existing Arm attacks -- Evict+ Time (\#41-\#42 in the Figure), 
Prime+Probe (\#43-\#44 in the Figure), Flush+Reload\footnote{
Our Flush+Reload benchmarks test for a stronger variant of the Evict+Reload
vulnerability shown in~\cite{green2017autolock,lipp2016armageddon}.
} (\#5-\#8 in the Figure), and 
Flush+Flush (\#47-\#50 in the Figure) -- all indeed show up as effective vulnerabilities 
for 30 out of the \dnum \, mobile devices tested.
This confirms that our benchmarks can cover existing work.
Note that the 5 types of vulnerabilities explored by prior work, e.g., the Evict+ Time, etc.,
can be realized using more than one vulnerability from the 88 types, thus prior work
covers 12 types, leaving 76 types not considered, for the total of 88 vulnerabilities that are possible.

\subsection{Summary of Vulnerability Trends}

To summarize, the patterns of the vulnerabilities uncovered
thanks to the systematic benchmarking on \dnum \, devices are:

\begin{itemize}[noitemsep]

  \item Microarchitectural features: performance increasing features such as the store buffer can degrade security,
  while features such as the snoop control unit can be helpful, indicating that security and performance are not always at odds with each other, and some features can help both.
  \item Heterogeneous cache size: 
  larger coherence timing for accesses involving cores in different clusters, compared to within same cluster,
  may lead to more vulnerabilities being effective.
  \item Core frequency: larger core frequency generally correlates with more vulnerabilities.
  \item WB and MSHR sizes: WB size does not impact security, while larger MSHR may allow more vulnerabilities to be effective.
  \item Vulnerability type effectiveness: relations of number of effective vulnerabilities showed  are: $AO$-Type \textgreater\, $SA$-Type \textgreater\, $SO$-Type; 
  meanwhile, $I$-Type and $E$-Type vulnerabilities are similarly effective on the tested devices.
  \item Tested device results: relations of number of effective vulnerabilities showed are:  Kryo 585  \textgreater\, Kryo 385 $\approx$ Kryo 360 \textgreater\,  
  Core A53 \textgreater\, Kryo 280 \textgreater\,   Kryo 260.
  
\end{itemize}

\section{Sensitivity Testing of Benchmarks}
\label{sec:sensitivity}

To understand how the benchmarks are affected by possible misconfigurations,
we performed a number of sensitivity tests.
In addition to evaluating how the benchmarks behave,
the sensitivity study allows us to understand how knowledge
(or lack of knowledge) of the correct cache configuration affects the attacker's ability
to attack the system.

\subsection{Analysis of Sensitivity Testing}

The most important cache parameters for sensitivity tests are: {\em associativity}, {\em line size}, and {\em total cache size}.
We use $asso_d$, $line_d$, and $tot_d$ to respectively denote the value of the parameters of the actual target device.
Meanwhile, $asso_b$, $line_b$, and $tot_b$ denote the cache parameters used by the benchmarks.
The parameters used in the tests are varied and are different from the actual, correct parameters
to test the sensitivity of the results to misconfiguration.
As we show, setting the configuration incorrectly in the benchmarks changes the mapping of the addresses 
used by the benchmarks, and influences the number of
vulnerabilities judged to be effective on a device.

We implement the sensitivity tests in the following way.
A large array is maintained to locate three different candidates of the secret value ($a$, $a_{alias}$, or $NIB$). 
We consider two addresses that only differ in the low $log_2(line_b)$ bits to belong to the same cache line, 
and two addresses that are a distance of $C\times tot_b/asso_b$ ($C$ is a integer) apart to map to the same cache set.
For each step, we access $asso_b$ number of addresses for each cache set to occupy or cause collision in the whole cache set.
To increase the signal to noise ratio in our measurements,
$rep$ cache sets are accessed in each of the steps of a benchmark (in our setting this number is 8).

When $asso_b$, $line_b$, or $tot_b$ deviates from $asso_d$, $line_d$, or $tot_d$, 
the following situations could happen:
\circled{1} the number of addresses being accessed in one cache set is less than $asso_d$, 
so interferences that should happen are not observed;
\circled{2} the addresses that should map to a target cache set actually map to several cache sets,
and contention in the target cache set might not happen or will become contention in several sets;
and \circled{3} the addresses that should map to different cache sets
actually map to the same cache set, introducing noise to the channel. 
We show later that the total number of attacks judged to be effective is less when an
incorrect configuration is used -- however, there are still attacks that are effectively independent of the configuration setting.

In the following, we denote one L1 cache hit timing as $t_{L1}$ and one L2 cache hit timing as $t_{L2}$.
When the configuration of the benchmark is correct, if the secret maps to the same cache set 
as some known address that was accessed, $t_{L2}$ will be observed, while if they are not mapped, $t_{L1}$ will be observed. 
In this case, timing observations for mapped and unmapped cases are $asso_d\times t_{L2}$ and $asso_d \times t_{L1}$.

\subsubsection{Cache Associativity}

Associativity usually influences the number of accesses that map to a target cache set.
We distinguish two cases:

\begin{itemize}[noitemsep]
  \item \textbf{$asso_b<asso_d$}: 
  In this case,
  due to smaller number of ways accessed in each step, fewer evictions will occur (situation \circled{1}).
  If a data 
  address maps to the same set as the secret data, timing observation will be 
  $n\times t_{L2} + (asso_d-n)\times t_{L1}$  instead of $asso_d\times t_{L2}$.
  Here, $0<n<asso_b$. Due to the random replacement policy, only $n$ (not all $asso_b$) cache lines will be evicted.
  This will make the timing less distinguishable compared with the unmapped case, in which timing should be equal to $asso_d\times t_{L1}$.

  \item \textbf{$asso_b>asso_d$}: 
  When $tot_b=tot_d$, this setting will lead to accesses  
  that should map to one cache set actually mapping to several 
  cache sets (situation \circled{2}). This will result in measuring more than $rep$ 
  of cache sets for one step, which possibly introduces more noise.

\end{itemize}

\subsubsection{Cache Line Size}

Line size generally influences which cache set is chosen within an attack (benchmark) step.
Again, we distinguish two cases:

\begin{itemize}[noitemsep]
  \item \textbf{$line_b<line_d$}:
  In this setting, the accesses that should map to different cache sets in the benchmark actually map to the same cache set
  (situation \circled{3}). 
  This will lead to the result that the benchmark measures less than $rep$ cache sets effectively, 
  causing a reduced signal to noise ratio. 
  For example, when choosing $line_b=line_d/2$, then two addresses that differ in $line_b$ 
  will map to the same cache line instead of different lines in difference sets. 
  This results in having more L1 cache hits, from $asso_d\times t_{L2}$ to  $asso_d/2 \times t_{L2}+asso_d/2 \times t_{L1}$, 
  which makes it less distinguishable compared with unmapped case where timing is $asso_d \times t_{L1}$.
  
  \item \textbf{$line_b>line_d$}:
  In this setting, since we always access the first 64 bits in a cache line, 
  the addresses that should map to the same sets in the benchmark (with the incorrect configuration)
  still map to the same set (if the correct configuration was used). 
  However, when $line_b$ is larger or equal to $cache\_set/rep$\footnote{In the example of Section~\ref{subsec:sens_example}, 
  this number is equal to $128/8=16$.} times of $line_d$, 
  the address for $NIB$ in the benchmark will wrap back and map to the same cache set as $a$ and $a_{alias}$ (situation \circled{3}), causing a false negative result.

\end{itemize}

\subsubsection{Total Cache Size}

Cache size mainly influences the data addresses accessed in each step of an attack (benchmark).

\begin{itemize}[noitemsep]
  \item \textbf{$tot_b<tot_d$}:
  In this setting, the accesses that should map to one cache set in the benchmark 
  actually map to several cache sets (situation \circled{2}), because $tot_b/asso_d<tot_d/asso_d$. 
  This further causes the number of  data accesses in each set to be less than the 
  number of ways being accessed in the target cache set, i.e., $asso_d$ (situation \circled{1}).
  Thus, for the mapped case, it is equivalent to 
  observing $n\times t_{L2}$ timing instead of $asso_d\times t_{L2}$ timing for this cache set, 
  where $0 < n < tot_b/tot_d\times asso_d$ due to the random replacement policy.
  This could decrease the signal to noise ratio.
  
  \item \textbf{$tot_b>tot_d$}:
  Let $C'=tot_b/tot_d$. In most cases, $C'$ will be an integer, assuming a cache size (both $tot_b$ and $tot_d$) of $2^N$ bytes.
  In this setting, the cache addresses that are different by $tot_b/asso_d=C'\times tot_d/asso_d$ in the benchmark,
  will still map to a different cache set in target device\footnote{When $C'$ is not an integer, e.g., $C'=1.5$, 
  then the address to set mapping will be different than the case when $tot_b=tot_d$, 
  which is equivalent to having addresses mapped to other cache set, 
  resulting in fewer number of addresses mapped to the target cache set (situation \circled{1}).}.
  Further, if $C'$ is too large, this will cause unexpected system noise 
  if prefetching, copy-on-write, etc., functions are enabled in the device.

\end{itemize}

\subsubsection{Analysis by Vulnerabilities Types}

For $AO$-Type and $SA$-Type Vulnerabilities,
the timing observation for $V_u=a$ is different from $V_u=a_{alias}$ or $V_u=NIB$.
In these types of vulnerabilities, the attack does not rely on the interference between different cache lines in a cache set. 
How the addresses map to the cache set does not affect the result, 
and the cache configurations will not influence the effectiveness of the 
vulnerabilities.
Also, these types usually rely on relatively larger timing differences, so the signal to noise ratio is large.

$SO$-Type 
vulnerabilities usually derive the $V_u$ information by observing evictions of the originally
accessed data in a prior attack step.
For $SO$-Type vulnerabilities, we need to access all the $asso_d$ ways to prime the whole cache set
in order to observe the timing difference, therefore, $SO$-Type 
vulnerabilities will actually be
influenced by the setting of parameters including 
{\em associativity}, {\em line size}, and {\em total cache size}.

\subsubsection{Summary}

Based on the above, we make three observations about the configurations' impact on the benchmarks and the corresponding attacks and how easy they are to perform:

\begin{itemize}[noitemsep]
  \item[1.] Attackers can still attack the system even when they are uncertain about the cache configuration.
  This is especially true for $AO$-Type or $SA$-Type attacks since they are not impacted much by the (mis) configuration.
  \item[2.] Most of the differences are due to 
  $SO$-Type attacks, which do not work well when incorrect setting is selected.
  \item[3.] Setting correct configurations causes 
  more vulnerabilities to be judged effective for a device. Incorrect settings can
  cause an underestimation of the total number of vulnerabilities.

\end{itemize}

\subsection{Evaluation of Sensitivity Testing} 
\label{subsec:sens_example}

We tested a wide range of devices and found similar trends among the results.
Here we give results for one example phone, Google Pixel 2, to show how the sensitivity test is implemented and evaluated.

The L1 cache configuration of small core of Google Pixel 2 is 32KB, 4-way 
set-associative with line size to be 64B.
We test this configuration by changing one of the three parameters (associativity, line size or cache size), 
while keeping the other two the same to avoid interference between different parameters.
The different configuration values we choose in our evaluation are listed in Table~\ref{tbl:sensitivity_test}.

\begin{figure}[t!]
  \label{fig:sensitivity_test}
        \centering
        \includegraphics[height=1.4in]{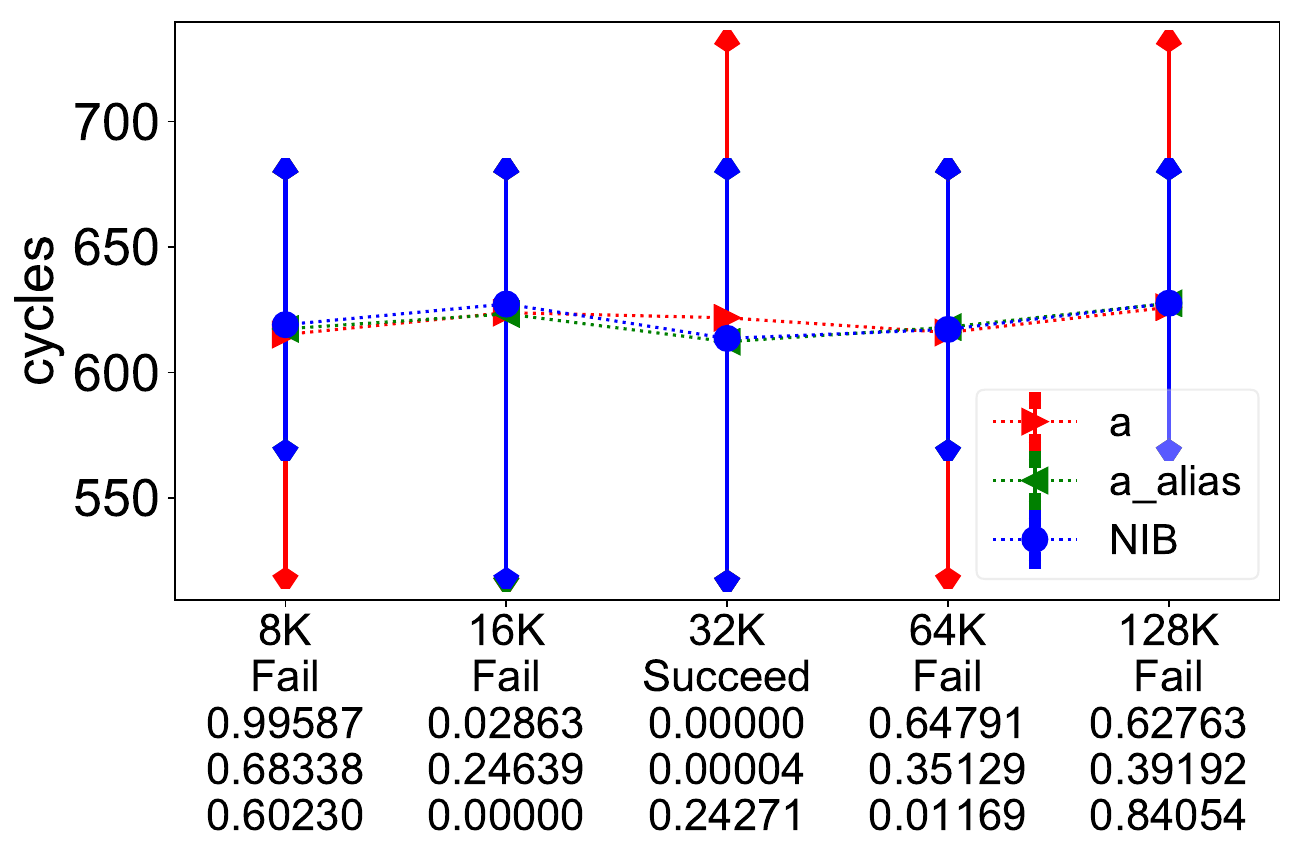}
    \caption{Timing histogram of a vulnerability case when changing the cache size. 
    The error bar shows the range of timing distribution and the dot shows the average timing cycles.
    ``Succeed'' under the configuration means the vulnerability is effective while ``Fail'' means not.
    Three values under ``Succeed'' or ``Fail'' are the {\em pvalue} for each two timing 
    distributions out of three. If it is smaller than 0.00049, we judge the two timing distributions to be differentiable, otherwise not.
    }
    \label{fig:sensitivity_test}
\end{figure}

\begin{table}[t!]
\small{
\caption{\small Configuration test results for cache associativity, line size and cache size of Google Pixel 2. 
Black bold numbers show the largest effective number of vulnerabilities for each category.
Middle column shows the correct configuration values for this device, other columns show smaller (left side) and bigger (left side)
values that were tested for each parameter of the cache.}
  \label{tbl:sensitivity_test}
  \begin{center}
  \footnotesize
    \begin{tabular}{|C{0.8cm}|C{1.85cm}|C{0.55cm}|C{0.55cm}||C{0.55cm}||C{0.55cm}|C{0.55cm}|}
    \hline
     Config. & & \multicolumn{5}{c|}{Effective Vul. Num. for Diff. Config.} \\ \hline \hline

     \multirow{3}{3.5em}{Assoc- iativity} & $asso_b$ Value & 1 & 2 & 4 & 8 & 16\\ \cline{2-7}
      & Total Vul. Num.  & 75 & 78  & \textbf{82}  & 75  & 75  \\ \cline{2-7}
       & $SO$-Type Num.  & 17 & 17  & \textbf{20}  & 13  & 12  \\ \hline
      \multirow{3}{3.5em}{Line Size} & $line_b$ Value & 16 & 32 & 64 & 128 & 256\\ \cline{2-7}
      & Total Vul. Num.  & 77 & 75  & \textbf{82}  & 80  & 79  \\ \cline{2-7}
       & $SO$-Type Num.  & 14 & 12  & \textbf{18} & 17  & 17  \\ \hline
      \multirow{3}{3.5em}{Cache Size} & $tot_b$ Value & 8192 & 16384 & 32768 & 65536 & 98304\\ \cline{2-7}
      & Total Vul. Num.  & 79 & 77  & \textbf{82}  & 79  & 77  \\ \cline{2-7}
       & $SO$-Type Num.  & 16 & 15  & \textbf{20}  & 16  & 14  \\ \hline
     
     \end{tabular}
  \end{center} }
\end{table}

In the example test case shown in Figure~\ref{fig:sensitivity_test}, timing distribution differences 
between three candidates are larger for the correct configuration, compared to the wrong configurations. 
The vulnerability is effective under the correct configuration while it fails for the incorrect configuration.

As shown in Table~\ref{tbl:sensitivity_test}, we found that
differences between the number of correct configuration and incorrect 
configuration for all effective vulnerabilities and $SO$-Type only effective vulnerabilities are roughly the same. For example,
when changing the associativity, difference of all effective vulnerability numbers 
between 4 (82) and 8 (75) is 7, which is the same as difference of $SO$-Type numbers (between 4 (20) and 8 (13)).
This also shows that wrong configurations will still lead to $AO$-Type and $SA$-Type vulnerabilities to be
effective even if the configuration is wrong.

As shown in Table~\ref{tbl:sensitivity_test} as well,
attacks are most effective under the correct configuration.
When setting the wrong value for either one of the three cache configurations, 
the number of vulnerabilities that are effective decreases.
On the other hand, this shows that hiding the cache architecture information or giving wrong configurations 
on-purpose is not a reliable defense.

\section{Evaluation of Secure Caches}
\label{sec:gem5_cache}

As shown in the previous sections,
current commercial Arm architectures are 
indeed vulnerable to most of the attack types.
A potential defense are secure caches.
To help understand if existing secure cache designs could help defend the attacks in Arm processors,
we implemented and evaluated the Partition-Locked (PL)~\cite{wang2007new} and Random Fill (RF)~\cite{liu2014random} caches 
together with our benchmarks in the {\tt gem5} simulator.
We show that they can defend many of the attacks, but we also uncover new vulnerabilities in the secure cache designs.
In this section, we focus on the security analysis of the secure cache designs. 
Performance evaluations
of PL cache and RF cache can be found in~\cite{wang2007new} and~\cite{liu2014random}, where reasonable overhead is shown.

\subsection{PL Cache Design and Implementation}

\begin{figure}[t]
\centering
\includegraphics[width=0.4\textwidth]{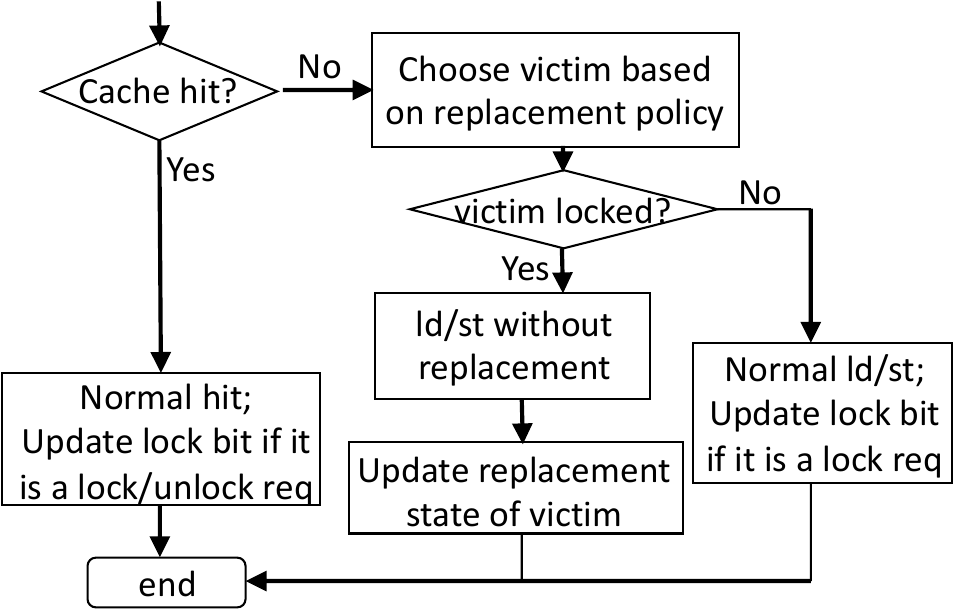}
 \caption{\small PL cache replacement logic flow-chart, as proposed in~\cite{wang2007new}. }
 \label{fig:PL_cache}
\end{figure}

Cache replacement is considered as the root cause of many cache side-channel attacks, 
and partitioned caches were proposed to prevent the victim's cache line from being evicted by the attacker.
PL cache~\cite{wang2007new} is a flexible cache partitioning design, where the victim can choose cache lines to be partitioned.  
In the PL cache, each cache line is extended with a lock bit to indicate if the line is locked in the cache.
When a cache line is locked, the line will not be evicted by any cache replacement until it is unlocked.
Figure~\ref{fig:PL_cache} shows the replacement logic of the PL cache. 
If a locked cache line is selected to be evicted, the eviction will not happen, and the incoming cache line will be handled uncached.
If the victim locks the secret-related address properly and the cache is big enough to hold all the locked cache lines, 
the PL cache is secure against all types of timing-based vulnerabilities, because the secret-related address will always be in the cache.

To evaluate the PL cache against different vulnerabilities, we implement it in the L1 data cache and 
add new instructions to lock (and unlock) cache lines in the {\tt gem5} simulator. 
The evaluation in {\tt gem5} is run in SE mode using a single O3CPU core, 
where each benchmark has an additional {\tt lock} step for locking the victim's cache line.

\begin{figure}[t!]
\centering
\includegraphics[width=0.42\textwidth]{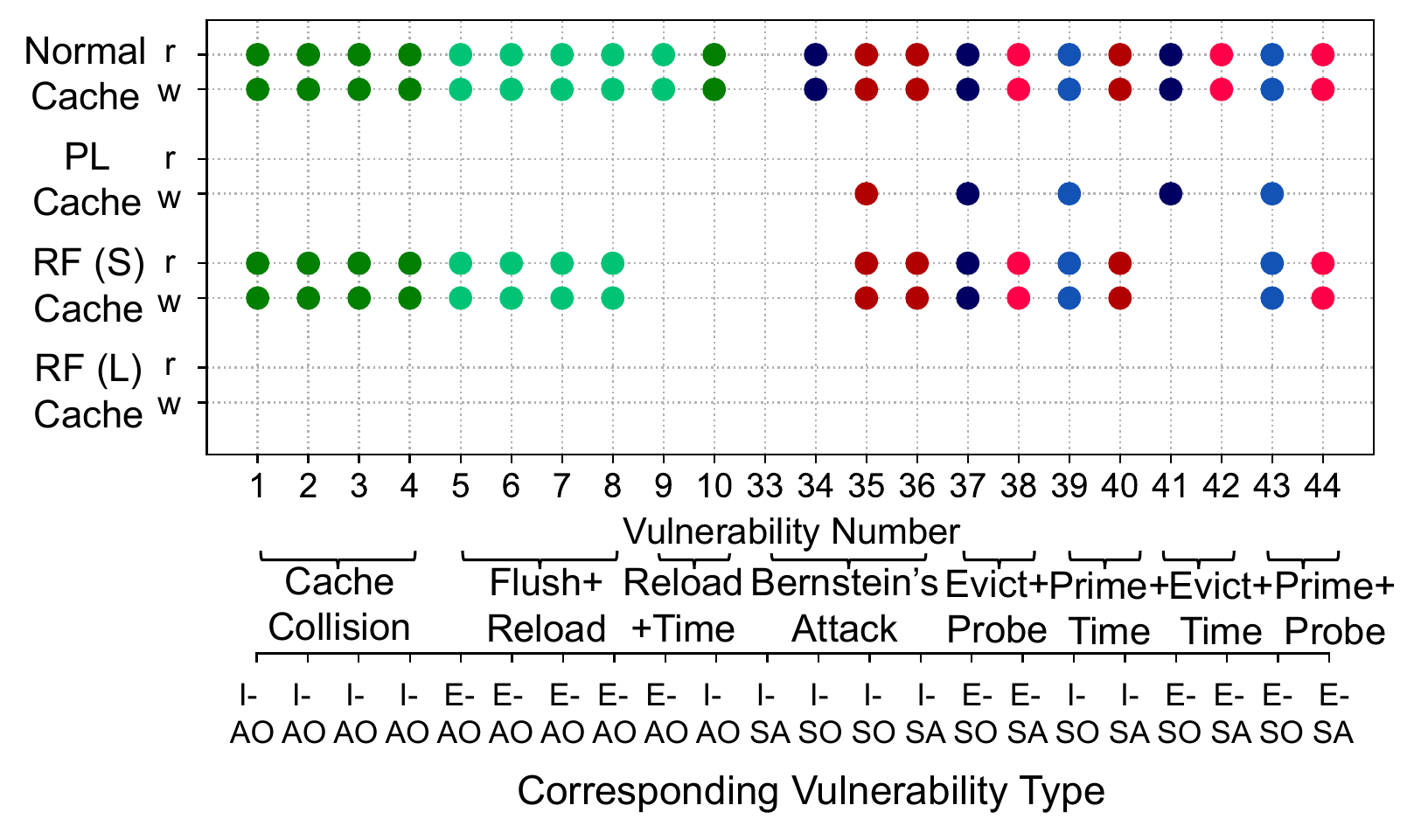} 
\caption{Evaluation results of security benchmarks on PL cache, RF cache, and a normal set-associative cache, for comparison. Solid dots, half solid dots or empty dot mean all of the, part of the, or no vulnerability cases are vulnerable to the cache, respectively.}
\label{fig:norm_se_rf}
\end{figure}

\subsection{Security Evaluation of the PL Cache}

Figure~\ref{fig:norm_se_rf} shows the results of evaluation of the PL cache (and the RF cache, as well as the baseline set-associative cache). 
For the PL cache, $AO$-Type vulnerabilities such as  Flush+ Reload fail, because the sensitive data is locked in the cache, and cannot be evicted by the benchmark
steps that simulate the attacker.
Without locking, 
a normal cache is vulnerable to these attacks, as shown in Figure~\ref{fig:norm_se_rf}.

For  $SO$-Type or $SA$-Type vulnerabilities such as Bernstein's attack,
theoretically the PL cache should prevent all of them as well.
However, from the experimental results we find that when the steps are implemented using write (store), some of the attacks will still be successful 
in the PL cache.
This is mainly due to the write buffer structure, which is not considered in original design of the PL cache~\cite{wang2007new}.
These attack strategies all require conflicts of known and unknown secret cache lines. 
Although being 
locked before the attack runs, the secret cache lines may be further brought into the write buffer due to a write access 
and then leave the cache structure to ``bypass'' the locking features, making the attack successful.
On the other hand, without the influence of the write buffer, we find that the attack cases that have 
all three steps to be non-write accesses to be always prevented on PL cache, as expected.
The vulnerabilities leveraging the cache coherence states and multiple cores were not considered in original PL cache design,
but can be tested in future. 

The PL cache evaluation highlights the need for systematic security evaluation using benchmarks.
Thanks to the approach, 
the original PL cache design is found to have a new write-based attack.
More importantly, our benchmarks can be useful for designing future secure caches and testing them in $gem5$.

\subsection{RF Cache Design and Implementation}

\begin{figure}[t]
\centering
\includegraphics[width=0.32\textwidth]{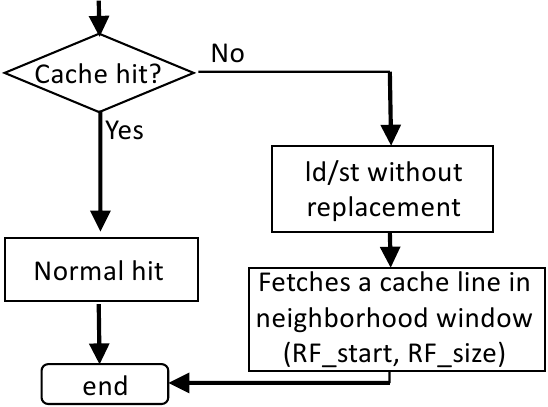}
 \caption{\small RF cache replacement logic flow-chart, as proposed in~\cite{liu2014random}. }
 \label{fig:RF_cache}
\end{figure}

To prevent interference caused by cache replacement, Random Fill (RF) cache~\cite{liu2014random} 
has been proposed to de-correlate the cache fill that causes the cache replacement and the victim's cache 
access. On a cache miss, the missing cache line will be handled without being fetched in the cache, 
instead a cache line in the neighborhood window $[addr-RF\_start, addr-RF\_start+RF\_size]$ will be fetched, as shown in 
Figure~\ref{fig:RF_cache}. In this way, the memory access pattern is de-correlated from the cache lines fetched 
in the cache. Since fetching cache lines in the neighborhood window may still carry information about 
the original $addr$, the security of RF cache depends on the parameters $RF\_start$ and $RF\_size$.

We implement the RF cache in the L1 data cache, as suggested by the work~\cite{liu2014random}. 
Note that here the cache line will still be fetched into L2 cache, but vulnerabilities targeting the L1 
cache should be defended. Parameters $RF\_start$ and $RF\_size$ can be configured in {\tt gem5}.
The benchmark suite for evaluation is identical to the normal three-step benchmarks, no additional step is required for the RF cache,
e.g., no special locking step is needed.

\subsection{Security Evaluation of the RF Cache}

RF cache can potentially defend all attacks because the victim's access to a secret address will not 
cause the corresponding cache line to be fetched into cache, but a random cache line in a neighborhood window 
will be fetched instead. However, fetching a cache line in the neighborhood window still can transfer information about the victim's cache access. 
We tested two different RF cache configurations, one with small neighborhood window (5 cache lines) and one with large neighborhood window (128 cache lines\footnote{There are 128 cache sets in the evaluated L1 cache.}). 

To reduce noise in the tests, the benchmarks test 8 contiguous cache lines and measure the total timing. 
When the neighborhood window of the RF cache is small, the cache line fetched into the cache will be 
not far from the address being accessed, and can still be observed by the third step of the benchmark 
with a high probability. As shown in Figure~\ref{fig:norm_se_rf}, for a small neighborhood window (S), 
a number of vulnerabilities are still effective, such as Flush+Reload and Prime+Probe.

For a large neighborhood window (L), no effective vulnerabilities are detected by the benchmark. 
For $SO$-Type vulnerabilities, the large neighborhood window de-correlates the memory access and the cache set to be accessed, 
so that the vulnerabilities can be prevented.
For $AO$-Type vulnerabilities, the channel capacity of the cache side channel decreases with 
the window size due to the reduced probability of the desired cache line being fetched into cache, 
as analyzed in~\cite{liu2014random}. The neighborhood window of 128 cache lines is enough to mitigate the channel in our setting where there are 128 cache sets.

The evaluation of the RF cache shows how the benchmark suite can be used to help choose the design parameter,
and the benchmark can quickly evaluate the design prototypes.

\subsection{Security Evaluation of Other Secure Caches}

CEASER~\cite{qureshi2018ceaser} is able to mitigate conflict-based
LLC timing–based side-channel attacks using address
encryption and dynamic remapping. The CEASER cache does
not differentiate whom the address belongs to and whether
the address is security critical. When a memory access tries
to modify the cache state, the address will first be encrypted
using a Low-Latency BlockCipher (LLBC)~\cite{borghoff2012low}, which not
only randomizes the cache set it maps to, but also scatters the
original, possibly ordered, and location-intensive addresses
to different cache sets, decreasing the probability of conflict
misses. 
The encryption key
will be periodically changed to avoid key reconstruction.
CEASER-S~\cite{qureshi2019new}
allows CEASER to 
divide the cache ways into multiple partitions of all the cache ways and allows the
line to be mapped to a different set in each partition
via principles of skewing.
The modified ``skew'' idea of CEASER-S cache assigns each partition 
a different multiple instance
of CEASER to determine the set mappings to strengthen the random mapping.
These two caches, focusing on randomizing cache set mapping, 
targets $SO$-type or $SA$-type attacks and cannot prevent $AO$-type
vulnerabilities.

ScatterCache~\cite{werner2019scattercache}
uses cache set randomization to
prevent timing-based attacks. It builds upon two ideas. First,
a mapping function is used to translate memory addresses
and process information to cache set indices. The mapping
is different for each program or security domain. Second,
the mapping function also calculates a different index for
each cache way. The mapping function can be keyed hash or
keyed permutation derivation function -- a different key is
used for each application or security domain resulting
in a different mapping from addresses to cache sets.
Software (e.g., the operating system) is responsible for
managing the security domains and process IDs, which are
used to differentiate the software processes and assign them with 
different keys for the mapping. As hardware extension,
a cryptographic primitive such as hashing and an index
decoder for each scattered cache way is added. 
ScatterCache
is able to prevent 
$SO$-type or $SA$-type vulnerabilities by assigning a different index 
for each cache way and security domain.
It encrypts both the cache address and process ID when
mapping into the cache index,
therefore, ScatterCache is able to prevent 
$E$-$AO$-type vulnerabilities 
such as
Flush+Reload, but not $I$-$AO$-type vulnerabilities such as Cache Collision vulnerabilities.

Time-Predictable Secure Cache (TSCache)~\cite{trilla2018cache}
relies on random placement to exhibit randomized
execution times. To achieve side-channel attack robustness, random placement
must also decouple cache interference of the attacker from
the victim. Memory addresses from victim and attacker’s
processes must not contend systematically in the same cache set.
Instead, each memory address from each process must be randomly
and independently placed in a set, thus randomizing interference.
This is achieved by operating the address
(tag and index bits) together with a random number called random seed.
Each task is forced to have a different seed so that conflicts between attacker's
and victim's cache lines are random and independent across runs,
thus defeating any contention-based attacks.
The same seed is given to
allow the communication between runnables of a
given software components of an application via shared memory.
TSCache exploits random placement to de-correlate set mapping with the corresponding
address index bits. Therefore, it can be used to prevent $SO$-type or $SA$-type vulnerabilities 
but may not be able to prevent $AO$-type vulnerabilities.

\section{Conclusion}
\label{sec:conclusion}

This paper presented for the first time a large-scale evaluation of \dnum \, Arm devices against 88 types
of vulnerabilities.  In total, three different cloud platforms were leveraged for the evaluation, and {\tt gem5} was used for
further analysis of certain microarchitectural features.
Based on the evaluation results, the work uncovered a number of components of the microarchitectual design that 
influence the effectiveness of different types of the vulnerabilities.
Further, sensitivity
tests were used to understand impacts of possible misconfiguration on the outcome of the benchmarks,
and also showed that even with uncertain cache configuration, number of attack types can be successful.
To help defend the attacks,
the PL and RF secure caches were implemented and evaluated on {\tt gem5}.
Based on the benchmarking results of the secure caches,
a new attack on PL cache, and possible issues due to small window size in the RF cache were uncovered.

\ifCLASSOPTIONcompsoc
  \section*{Acknowledgments}
\else
  \section*{Acknowledgment}
\fi

This work was supported in part by NSF grants 1651945 and 1813797, and through SRC task 2488.001.
The authors would like to acknowledge Amazon Web Services for cloud research
credits used for some of the testing.

\ifCLASSOPTIONcaptionsoff
  \newpage
\fi

\section*{Appendix}

\begin{table*}[ht!]
\small
\caption{\small {\em Pvalue} evaluation results for different secure caches.}
  \label{tbl_p_value}
  \begin{center}
  \fontsize{6pt}{7.3pt}\selectfont
    \begin{tabular}{| C{0.26cm} | C{0.26cm} | C{0.26cm} | C{0.26cm} || C{0.26cm} | C{0.26cm} | C{0.26cm} || C{0.26cm} | C{0.26cm} | C{0.26cm} || C{0.26cm} | C{0.26cm} | C{0.26cm} ||  C{0.26cm} |C{0.26cm} | C{0.26cm} | C{0.26cm} || C{0.26cm} | C{0.26cm} | C{0.26cm} || C{0.26cm} | C{0.26cm} | C{0.26cm} || C{0.26cm} | C{0.26cm} | C{0.26cm} |} 
    \hline
    \multicolumn{26}{|c|}{Normal Cache} \\ \hline
     Vul & \multicolumn{3}{c||}{F-R-R} & \multicolumn{3}{c||}{F-R-W} & \multicolumn{3}{c||}{F-W-R}& \multicolumn{3}{c||}{F-W-W} & Vul
      & \multicolumn{3}{c||}{F-R-R} & \multicolumn{3}{c||}{F-R-W} & \multicolumn{3}{c||}{F-W-R}& \multicolumn{3}{c|}{F-W-W} \\ \hline 
     
1 & 0.00 & \cellcolor{blue!25}0.93 & 0.00 & 0.00 & \cellcolor{blue!25}0.64 & 0.00 & 0.00 & \cellcolor{blue!25}0.79 & 0.00 & 0.00 & \cellcolor{blue!25}0.67 & 0.00 & 2 & 0.00 & \cellcolor{blue!25}0.93 & 0.00 & 0.00 & \cellcolor{blue!25}0.64 & 0.00 & 0.00 & \cellcolor{blue!25}0.79 & 0.00 & 0.00 & 0.00 & \cellcolor{blue!25}0.24 \\ \hline
3 & 0.00 & \cellcolor{blue!25}0.95 & 0.00 & 0.00 & \cellcolor{blue!25}0.29 & 0.00 & 0.00 & \cellcolor{blue!25}0.04 & 0.00 & 0.00 & \cellcolor{blue!25}0.43 & 0.00 & 4 & 0.00 & \cellcolor{blue!25}0.95 & 0.00 & 0.00 & \cellcolor{blue!25}0.29 & 0.00 & 0.00 & \cellcolor{blue!25}0.04 & 0.00 & 0.00 & \cellcolor{blue!25}0.32 & 0.00 \\ \hline
5 & 0.00 & \cellcolor{blue!25}0.95 & 0.00 & 0.00 & \cellcolor{blue!25}0.29 & 0.00 & 0.00 & \cellcolor{blue!25}0.04 & 0.00 & 0.00 & \cellcolor{blue!25}0.43 & 0.00 & 6 & 0.00 & \cellcolor{blue!25}0.95 & 0.00 & 0.00 & \cellcolor{blue!25}0.29 & 0.00 & 0.00 & \cellcolor{blue!25}0.04 & 0.00 & 0.00 & \cellcolor{blue!25}0.32 & 0.00 \\ \hline
7 & 0.00 & \cellcolor{blue!25}0.93 & 0.00 & 0.00 & \cellcolor{blue!25}0.64 & 0.00 & 0.00 & \cellcolor{blue!25}0.79 & 0.00 & 0.00 & \cellcolor{blue!25}0.67 & 0.00 & 8 & 0.00 & \cellcolor{blue!25}0.93 & 0.00 & 0.00 & \cellcolor{blue!25}0.64 & 0.00 & 0.00 & \cellcolor{blue!25}0.79 & 0.00 & 0.00 & 0.00 & \cellcolor{blue!25}0.24 \\ \hline
9 & 0.00 & \cellcolor{blue!25}0.13 & 0.00 & 0.00 & \cellcolor{blue!25}0.01 & 0.00 & 0.00 & 0.00 & 0.00 & 0.00 & 0.00 & 0.00 & 10 & 0.00 & \cellcolor{blue!25}0.13 & 0.00 & 0.00 & \cellcolor{blue!25}0.01 & 0.00 & 0.00 & 0.00 & 0.00 & 0.00 & 0.00 & 0.00 \\ \hline
Vul & \multicolumn{3}{c||}{R-R-R} & \multicolumn{3}{c||}{R-R-W} & \multicolumn{3}{c||}{R-W-R}& \multicolumn{3}{c||}{R-W-W} & Vul
& \multicolumn{3}{c||}{W-R-R} & \multicolumn{3}{c||}{W-R-W} & \multicolumn{3}{c||}{W-W-R}& \multicolumn{3}{c|}{W-W-W} \\ \hline 
33 & 0.00 & 0.00 & 0.00 & 0.00 & 0.00 & 0.00 & 0.00 & 0.00 & 0.00 & 0.00 & 0.00 & 0.00 & 33 & 0.00 & 0.00 & 0.00 & 0.00 & 0.00 & 0.00 & 0.00 & 0.00 & 0.00 & 0.00 & 0.00 & 0.00 \\ \hline
34 & \cellcolor{blue!25}0.61 & 0.00 & 0.00 & \cellcolor{blue!25}0.62 & 0.00 & 0.00 & \cellcolor{blue!25}0.98 & 0.00 & 0.00 & \cellcolor{blue!25}0.97 & 0.00 & 0.00 & 34 & \cellcolor{blue!25}0.59 & 0.00 & 0.00 & \cellcolor{blue!25}0.64 & 0.00 & 0.00 & \cellcolor{blue!25}0.97 & 0.00 & 0.00 & \cellcolor{blue!25}0.94 & 0.00 & 0.00 \\ \hline
35 & \cellcolor{blue!25}0.17 & 0.00 & 0.00 & \cellcolor{blue!25}0.37 & 0.00 & 0.00 & \cellcolor{blue!25}0.32 & 0.00 & 0.00 & \cellcolor{blue!25}0.32 & 0.00 & 0.00 & 35 & \cellcolor{blue!25}0.19 & 0.00 & 0.00 & \cellcolor{blue!25}0.30 & 0.00 & 0.00 & \cellcolor{blue!25}0.32 & 0.00 & 0.00 & \cellcolor{blue!25}0.29 & 0.00 & 0.00 \\ \hline
36 & 0.00 & \cellcolor{blue!25}0.32 & 0.00 & 0.00 & \cellcolor{blue!25}0.41 & 0.00 & 0.00 & \cellcolor{blue!25}0.05 & 0.00 & 0.00 & \cellcolor{blue!25}0.27 & 0.00 & 36 & 0.00 & \cellcolor{blue!25}0.30 & 0.00 & 0.00 & \cellcolor{blue!25}0.37 & 0.00 & 0.00 & \cellcolor{blue!25}0.32 & 0.00 & 0.00 & \cellcolor{blue!25}0.34 & 0.00 \\ \hline
37 & \cellcolor{blue!25}0.17 & 0.00 & 0.00 & \cellcolor{blue!25}0.37 & 0.00 & 0.00 & \cellcolor{blue!25}0.32 & 0.00 & 0.00 & \cellcolor{blue!25}0.32 & 0.00 & 0.00 & 37 & \cellcolor{blue!25}0.19 & 0.00 & 0.00 & \cellcolor{blue!25}0.30 & 0.00 & 0.00 & \cellcolor{blue!25}0.32 & 0.00 & 0.00 & \cellcolor{blue!25}0.29 & 0.00 & 0.00 \\ \hline
38 & 0.00 & \cellcolor{blue!25}0.32 & 0.00 & 0.00 & \cellcolor{blue!25}0.41 & 0.00 & 0.00 & \cellcolor{blue!25}0.05 & 0.00 & 0.00 & \cellcolor{blue!25}0.27 & 0.00 & 38 & 0.00 & \cellcolor{blue!25}0.30 & 0.00 & 0.00 & \cellcolor{blue!25}0.37 & 0.00 & 0.00 & \cellcolor{blue!25}0.32 & 0.00 & 0.00 & \cellcolor{blue!25}0.34 & 0.00 \\ \hline
39 & \cellcolor{blue!25}0.17 & 0.00 & 0.00 & \cellcolor{blue!25}0.37 & 0.00 & 0.00 & \cellcolor{blue!25}0.32 & 0.00 & 0.00 & \cellcolor{blue!25}0.32 & 0.00 & 0.00 & 39 & \cellcolor{blue!25}0.19 & 0.00 & 0.00 & \cellcolor{blue!25}0.30 & 0.00 & 0.00 & \cellcolor{blue!25}0.32 & 0.00 & 0.00 & \cellcolor{blue!25}0.29 & 0.00 & 0.00 \\ \hline
40 & 0.00 & \cellcolor{blue!25}0.32 & 0.00 & 0.00 & \cellcolor{blue!25}0.41 & 0.00 & 0.00 & \cellcolor{blue!25}0.05 & 0.00 & 0.00 & \cellcolor{blue!25}0.27 & 0.00 & 40 & 0.00 & \cellcolor{blue!25}0.30 & 0.00 & 0.00 & \cellcolor{blue!25}0.37 & 0.00 & 0.00 & \cellcolor{blue!25}0.32 & 0.00 & 0.00 & \cellcolor{blue!25}0.34 & 0.00 \\ \hline
41 & \cellcolor{blue!25}0.61 & 0.00 & 0.00 & \cellcolor{blue!25}0.62 & 0.00 & 0.00 & \cellcolor{blue!25}0.98 & 0.00 & 0.00 & \cellcolor{blue!25}0.97 & 0.00 & 0.00 & 41 & \cellcolor{blue!25}0.59 & 0.00 & 0.00 & \cellcolor{blue!25}0.64 & 0.00 & 0.00 & \cellcolor{blue!25}0.97 & 0.00 & 0.00 & \cellcolor{blue!25}0.94 & 0.00 & 0.00 \\ \hline
42 & 0.00 & 0.00 & 0.00 & 0.00 & 0.00 & 0.00 & 0.00 & 0.00 & 0.00 & 0.00 & 0.00 & 0.00 & 42 & 0.00 & 0.00 & 0.00 & 0.00 & 0.00 & 0.00 & 0.00 & 0.00 & 0.00 & 0.00 & 0.00 & 0.00 \\ \hline
43 & \cellcolor{blue!25}0.17 & 0.00 & 0.00 & \cellcolor{blue!25}0.37 & 0.00 & 0.00 & \cellcolor{blue!25}0.32 & 0.00 & 0.00 & \cellcolor{blue!25}0.32 & 0.00 & 0.00 & 43 & \cellcolor{blue!25}0.19 & 0.00 & 0.00 & \cellcolor{blue!25}0.30 & 0.00 & 0.00 & \cellcolor{blue!25}0.32 & 0.00 & 0.00 & \cellcolor{blue!25}0.29 & 0.00 & 0.00 \\ \hline
44 & 0.00 & \cellcolor{blue!25}0.32 & 0.00 & 0.00 & \cellcolor{blue!25}0.41 & 0.00 & 0.00 & \cellcolor{blue!25}0.05 & 0.00 & 0.00 & \cellcolor{blue!25}0.27 & 0.00 & 44 & 0.00 & \cellcolor{blue!25}0.30 & 0.00 & 0.00 & \cellcolor{blue!25}0.37 & 0.00 & 0.00 & \cellcolor{blue!25}0.32 & 0.00 & 0.00 & \cellcolor{blue!25}0.34 & 0.00 \\ \hline
\hline

\multicolumn{26}{|c|}{PL Cache} \\ \hline
Vul & \multicolumn{3}{c||}{F-R-R} & \multicolumn{3}{c||}{F-R-W} & \multicolumn{3}{c||}{F-W-R}& \multicolumn{3}{c||}{F-W-W} & Vul
& \multicolumn{3}{c||}{F-R-R} & \multicolumn{3}{c||}{F-R-W} & \multicolumn{3}{c||}{F-W-R}& \multicolumn{3}{c|}{F-W-W} \\ \hline
1 & \cellcolor{blue!25}0.58 & \cellcolor{blue!25}0.32 & 0.00 & \cellcolor{blue!25}0.69 & \cellcolor{blue!25}0.32 & 0.00 & \cellcolor{blue!25}0.74 & \cellcolor{blue!25}0.32 & 0.00 & \cellcolor{blue!25}0.72 & \cellcolor{blue!25}0.32 & 0.00 & 2 & \cellcolor{blue!25}0.58 & \cellcolor{blue!25}0.32 & 0.00 & \cellcolor{blue!25}0.69 & \cellcolor{blue!25}0.32 & 0.00 & \cellcolor{blue!25}0.74 & \cellcolor{blue!25}0.32 & 0.00 & 0.00 & 0.00 & \cellcolor{blue!25}0.52 \\ \hline
3 & \cellcolor{blue!25}0.71 & \cellcolor{blue!25}0.33 & 0.00 & \cellcolor{blue!25}0.86 & \cellcolor{blue!25}0.32 & 0.00 & \cellcolor{blue!25}0.98 & \cellcolor{blue!25}0.33 & 0.00 & \cellcolor{blue!25}0.90 & \cellcolor{blue!25}0.29 & 0.00 & 4 & \cellcolor{blue!25}0.71 & \cellcolor{blue!25}0.33 & 0.00 & \cellcolor{blue!25}0.86 & \cellcolor{blue!25}0.32 & 0.00 & \cellcolor{blue!25}0.98 & \cellcolor{blue!25}0.33 & 0.00 & \cellcolor{blue!25}0.83 & \cellcolor{blue!25}0.32 & 0.00 \\ \hline
5 & \cellcolor{blue!25}0.71 & \cellcolor{blue!25}0.33 & 0.00 & \cellcolor{blue!25}0.86 & \cellcolor{blue!25}0.32 & 0.00 & \cellcolor{blue!25}0.98 & \cellcolor{blue!25}0.33 & 0.00 & \cellcolor{blue!25}0.90 & \cellcolor{blue!25}0.29 & 0.00 & 6 & \cellcolor{blue!25}0.71 & \cellcolor{blue!25}0.33 & 0.00 & \cellcolor{blue!25}0.86 & \cellcolor{blue!25}0.32 & 0.00 & \cellcolor{blue!25}0.98 & \cellcolor{blue!25}0.33 & 0.00 & \cellcolor{blue!25}0.83 & \cellcolor{blue!25}0.32 & 0.00 \\ \hline
7 & \cellcolor{blue!25}0.58 & \cellcolor{blue!25}0.32 & 0.00 & \cellcolor{blue!25}0.69 & \cellcolor{blue!25}0.32 & 0.00 & \cellcolor{blue!25}0.74 & \cellcolor{blue!25}0.32 & 0.00 & \cellcolor{blue!25}0.72 & \cellcolor{blue!25}0.32 & 0.00 & 8 & \cellcolor{blue!25}0.58 & \cellcolor{blue!25}0.32 & 0.00 & \cellcolor{blue!25}0.69 & \cellcolor{blue!25}0.32 & 0.00 & \cellcolor{blue!25}0.74 & \cellcolor{blue!25}0.32 & 0.00 & 0.00 & 0.00 & \cellcolor{blue!25}0.52 \\ \hline
9 & \cellcolor{blue!25}0.59 & \cellcolor{blue!25}0.85 & \cellcolor{blue!25}0.44 & \cellcolor{blue!25}0.56 & \cellcolor{blue!25}0.85 & \cellcolor{blue!25}0.41 & \cellcolor{blue!25}0.59 & \cellcolor{blue!25}0.83 & \cellcolor{blue!25}0.43 & \cellcolor{blue!25}0.56 & \cellcolor{blue!25}0.84 & \cellcolor{blue!25}0.41 &   10 & \cellcolor{blue!25}0.59 & \cellcolor{blue!25}0.85 & \cellcolor{blue!25}0.44 & \cellcolor{blue!25}0.56 & \cellcolor{blue!25}0.85 & \cellcolor{blue!25}0.41 & \cellcolor{blue!25}0.59 & \cellcolor{blue!25}0.83 & \cellcolor{blue!25}0.43 & \cellcolor{blue!25}0.56 & \cellcolor{blue!25}0.84 & \cellcolor{blue!25}0.41 \\ \hline
Vul & \multicolumn{3}{c||}{R-R-R} & \multicolumn{3}{c||}{R-R-W} & \multicolumn{3}{c||}{R-W-R}& \multicolumn{3}{c||}{R-W-W} & Vul
& \multicolumn{3}{c||}{W-R-R} & \multicolumn{3}{c||}{W-R-W} & \multicolumn{3}{c||}{W-W-R}& \multicolumn{3}{c|}{W-W-W} \\ \hline
33 & \cellcolor{blue!25}0.68 & \cellcolor{blue!25}0.81 & \cellcolor{blue!25}0.50 & \cellcolor{blue!25}0.54 & \cellcolor{blue!25}0.85 & \cellcolor{blue!25}0.40 & \cellcolor{blue!25}0.69 & \cellcolor{blue!25}0.83 & \cellcolor{blue!25}0.53 & \cellcolor{blue!25}0.56 & \cellcolor{blue!25}0.85 & \cellcolor{blue!25}0.42 & 33 & \cellcolor{blue!25}0.59 & \cellcolor{blue!25}0.85 & \cellcolor{blue!25}0.44 & \cellcolor{blue!25}0.56 & \cellcolor{blue!25}0.85 & \cellcolor{blue!25}0.41 & \cellcolor{blue!25}0.59 & \cellcolor{blue!25}0.83 & \cellcolor{blue!25}0.43 & \cellcolor{blue!25}0.56 & \cellcolor{blue!25}0.84 & \cellcolor{blue!25}0.41 \\ \hline
34 & \cellcolor{blue!25}0.05 & \cellcolor{blue!25}0.70 & \cellcolor{blue!25}0.02 & \cellcolor{blue!25}0.70 & \cellcolor{blue!25}0.04 & \cellcolor{blue!25}0.02 & \cellcolor{blue!25}0.91 & \cellcolor{blue!25}0.67 & \cellcolor{blue!25}0.58 & \cellcolor{blue!25}0.84 & \cellcolor{blue!25}0.68 & \cellcolor{blue!25}0.54 & 34 & \cellcolor{blue!25}0.02 & \cellcolor{blue!25}0.63 & \cellcolor{blue!25}0.83 & \cellcolor{blue!25}0.72 & \cellcolor{blue!25}0.88 & \cellcolor{blue!25}0.93 & \cellcolor{blue!25}0.57 & \cellcolor{blue!25}0.50 & \cellcolor{blue!25}0.84 & \cellcolor{blue!25}0.61 & \cellcolor{blue!25}0.47 \\ \hline
35 & \cellcolor{blue!25}0.03 & \cellcolor{blue!25}0.16 & 0.00 & 0.00 & \cellcolor{blue!25}0.15 & 0.00 & \cellcolor{blue!25}0.37 & \cellcolor{blue!25}0.16 & 0.00 & \cellcolor{blue!25}0.41 & \cellcolor{blue!25}0.12 & 0.00 & 35 & \cellcolor{blue!25}0.01 & \cellcolor{blue!25}0.12 & 0.00 & 0.00 & \cellcolor{blue!25}0.17 & 0.00 & \cellcolor{blue!25}0.39 & \cellcolor{blue!25}0.11 & 0.00 & \cellcolor{blue!25}0.42 & \cellcolor{blue!25}0.12 & 0.00 \\ \hline
36 & \cellcolor{blue!25}0.67 & \cellcolor{blue!25}0.31 & 0.00 & \cellcolor{blue!25}0.73 & \cellcolor{blue!25}0.34 & 0.00 & \cellcolor{blue!25}0.79 & \cellcolor{blue!25}0.33 & 0.00 & \cellcolor{blue!25}0.79 & \cellcolor{blue!25}0.32 & 0.00 & 36 & \cellcolor{blue!25}0.74 & \cellcolor{blue!25}0.30 & 0.00 & \cellcolor{blue!25}0.76 & \cellcolor{blue!25}0.32 & 0.00 & \cellcolor{blue!25}0.83 & \cellcolor{blue!25}0.32 & 0.00 & \cellcolor{blue!25}0.79 & \cellcolor{blue!25}0.32 & 0.00 \\ \hline
37 & \cellcolor{blue!25}0.03 & \cellcolor{blue!25}0.16 & 0.00 & 0.00 & \cellcolor{blue!25}0.15 & 0.00 & \cellcolor{blue!25}0.37 & \cellcolor{blue!25}0.16 & 0.00 & \cellcolor{blue!25}0.41 & \cellcolor{blue!25}0.12 & 0.00 & 37 & \cellcolor{blue!25}0.01 & \cellcolor{blue!25}0.12 & 0.00 & 0.00 & \cellcolor{blue!25}0.17 & 0.00 & \cellcolor{blue!25}0.39 & \cellcolor{blue!25}0.11 & 0.00 & \cellcolor{blue!25}0.42 & \cellcolor{blue!25}0.12 & 0.00 \\ \hline
38 & \cellcolor{blue!25}0.67 & \cellcolor{blue!25}0.31 & 0.00 & \cellcolor{blue!25}0.73 & \cellcolor{blue!25}0.34 & 0.00 & \cellcolor{blue!25}0.79 & \cellcolor{blue!25}0.33 & 0.00 & \cellcolor{blue!25}0.79 & \cellcolor{blue!25}0.32 & 0.00 & 38 & \cellcolor{blue!25}0.74 & \cellcolor{blue!25}0.30 & 0.00 & \cellcolor{blue!25}0.76 & \cellcolor{blue!25}0.32 & 0.00 & \cellcolor{blue!25}0.83 & \cellcolor{blue!25}0.32 & 0.00 & \cellcolor{blue!25}0.79 & \cellcolor{blue!25}0.32 & 0.00 \\ \hline
39 & \cellcolor{blue!25}0.03 & \cellcolor{blue!25}0.16 & 0.00 & 0.00 & \cellcolor{blue!25}0.15 & 0.00 & \cellcolor{blue!25}0.37 & \cellcolor{blue!25}0.16 & 0.00 & \cellcolor{blue!25}0.41 & \cellcolor{blue!25}0.12 & 0.00 & 39 & \cellcolor{blue!25}0.01 & \cellcolor{blue!25}0.12 & 0.00 & 0.00 & \cellcolor{blue!25}0.17 & 0.00 & \cellcolor{blue!25}0.39 & \cellcolor{blue!25}0.11 & 0.00 & \cellcolor{blue!25}0.42 & \cellcolor{blue!25}0.12 & 0.00 \\ \hline
40 & \cellcolor{blue!25}0.67 & \cellcolor{blue!25}0.31 & 0.00 & \cellcolor{blue!25}0.73 & \cellcolor{blue!25}0.34 & 0.00 & \cellcolor{blue!25}0.79 & \cellcolor{blue!25}0.33 & 0.00 & \cellcolor{blue!25}0.79 & \cellcolor{blue!25}0.32 & 0.00 & 40 & \cellcolor{blue!25}0.74 & \cellcolor{blue!25}0.30 & 0.00 & \cellcolor{blue!25}0.76 & \cellcolor{blue!25}0.32 & 0.00 & \cellcolor{blue!25}0.83 & \cellcolor{blue!25}0.32 & 0.00 & \cellcolor{blue!25}0.79 & \cellcolor{blue!25}0.32 & 0.00 \\ \hline
41 & \cellcolor{blue!25}0.05 & \cellcolor{blue!25}0.70 & \cellcolor{blue!25}0.02 & 0.00 & \cellcolor{blue!25}0.73 & 0.00 & \cellcolor{blue!25}0.91 & \cellcolor{blue!25}0.67 & \cellcolor{blue!25}0.58 & \cellcolor{blue!25}0.84 & \cellcolor{blue!25}0.68 & \cellcolor{blue!25}0.54 & 41 & \cellcolor{blue!25}0.02 & \cellcolor{blue!25}0.63 & 0.00 & 0.00 & \cellcolor{blue!25}0.65 & 0.00 & \cellcolor{blue!25}0.93 & \cellcolor{blue!25}0.57 & \cellcolor{blue!25}0.50 & \cellcolor{blue!25}0.84 & \cellcolor{blue!25}0.61 & \cellcolor{blue!25}0.47 \\ \hline
42 & \cellcolor{blue!25}0.68 & \cellcolor{blue!25}0.81 & \cellcolor{blue!25}0.50 & \cellcolor{blue!25}0.54 & \cellcolor{blue!25}0.85 & \cellcolor{blue!25}0.40 & \cellcolor{blue!25}0.69 & \cellcolor{blue!25}0.83 & \cellcolor{blue!25}0.53 & \cellcolor{blue!25}0.56 & \cellcolor{blue!25}0.85 & \cellcolor{blue!25}0.42 & 42 & \cellcolor{blue!25}0.59 & \cellcolor{blue!25}0.85 & \cellcolor{blue!25}0.44 & \cellcolor{blue!25}0.56 & \cellcolor{blue!25}0.85 & \cellcolor{blue!25}0.41 & \cellcolor{blue!25}0.59 & \cellcolor{blue!25}0.83 & \cellcolor{blue!25}0.43 & \cellcolor{blue!25}0.56 & \cellcolor{blue!25}0.84 & \cellcolor{blue!25}0.41 \\ \hline
43 & \cellcolor{blue!25}0.03 & \cellcolor{blue!25}0.16 & 0.00 & 0.00 & \cellcolor{blue!25}0.15 & 0.00 & \cellcolor{blue!25}0.37 & \cellcolor{blue!25}0.16 & 0.00 & \cellcolor{blue!25}0.41 & \cellcolor{blue!25}0.12 & 0.00 & 43 & \cellcolor{blue!25}0.01 & \cellcolor{blue!25}0.12 & 0.00 & 0.00 & \cellcolor{blue!25}0.17 & 0.00 & \cellcolor{blue!25}0.39 & \cellcolor{blue!25}0.11 & 0.00 & \cellcolor{blue!25}0.42 & \cellcolor{blue!25}0.12 & 0.00 \\ \hline
44 & \cellcolor{blue!25}0.67 & \cellcolor{blue!25}0.31 & 0.00 & \cellcolor{blue!25}0.73 & \cellcolor{blue!25}0.34 & 0.00 & \cellcolor{blue!25}0.79 & \cellcolor{blue!25}0.33 & 0.00 & \cellcolor{blue!25}0.79 & \cellcolor{blue!25}0.32 & 0.00 & 44 & \cellcolor{blue!25}0.74 & \cellcolor{blue!25}0.30 & 0.00 & \cellcolor{blue!25}0.76 & \cellcolor{blue!25}0.32 & 0.00 & \cellcolor{blue!25}0.83 & \cellcolor{blue!25}0.32 & 0.00 & \cellcolor{blue!25}0.79 & \cellcolor{blue!25}0.32 & 0.00 \\ \hline    
\hline

\multicolumn{26}{|c|}{RF (S) Cache} \\ \hline
Vul & \multicolumn{3}{c||}{F-R-R} & \multicolumn{3}{c||}{F-R-W} & \multicolumn{3}{c||}{F-W-R}& \multicolumn{3}{c||}{F-W-W} & Vul
& \multicolumn{3}{c||}{F-R-R} & \multicolumn{3}{c||}{F-R-W} & \multicolumn{3}{c||}{F-W-R}& \multicolumn{3}{c|}{F-W-W} \\ \hline 
1 & 0.00 & 0.00 & 0.00 & 0.00 & 0.00 & 0.00 & 0.00 & 0.00 & 0.00 & 0.00 & 0.00 & 0.00 & 2 & 0.00 & 0.00 & 0.00 & 0.00 & 0.00 & 0.00 & 0.00 & 0.00 & 0.00 & 0.00 & 0.00 & 0.00 \\ \hline
3 & 0.00 & 0.00 & 0.00 & 0.00 & 0.00 & 0.00 & 0.00 & 0.00 & 0.00 & 0.00 & 0.00 & 0.00 & 4 & 0.00 & 0.00 & 0.00 & 0.00 & 0.00 & 0.00 & 0.00 & 0.00 & 0.00 & 0.00 & \cellcolor{blue!25}0.01 & 0.00 \\ \hline
5 & 0.00 & 0.00 & 0.00 & 0.00 & 0.00 & 0.00 & 0.00 & 0.00 & 0.00 & 0.00 & 0.00 & 0.00 & 6 & 0.00 & 0.00 & 0.00 & 0.00 & 0.00 & 0.00 & 0.00 & 0.00 & 0.00 & 0.00 & \cellcolor{blue!25}0.01 & 0.00 \\ \hline
7 & 0.00 & 0.00 & 0.00 & 0.00 & 0.00 & 0.00 & 0.00 & 0.00 & 0.00 & 0.00 & 0.00 & 0.00 & 8 & 0.00 & 0.00 & 0.00 & 0.00 & 0.00 & 0.00 & 0.00 & 0.00 & 0.00 & 0.00 & 0.00 & 0.00 \\ \hline
9 & \cellcolor{blue!25}0.41 & \cellcolor{blue!25}0.11 & \cellcolor{blue!25}0.11 & \cellcolor{blue!25}0.42 & \cellcolor{blue!25}0.16 & \cellcolor{blue!25}0.15 & \cellcolor{blue!25}0.05 & \cellcolor{blue!25}0.02 & \cellcolor{blue!25}0.25 & \cellcolor{blue!25}0.18 & \cellcolor{blue!25}0.04 & \cellcolor{blue!25}0.18 & 10 & \cellcolor{blue!25}0.10 & \cellcolor{blue!25}0.04 & \cellcolor{blue!25}0.34 & \cellcolor{blue!25}0.11 & \cellcolor{blue!25}0.05 & \cellcolor{blue!25}0.38 & \cellcolor{blue!25}0.06 & \cellcolor{blue!25}0.03 & \cellcolor{blue!25}0.49 & \cellcolor{blue!25}0.09 & \cellcolor{blue!25}0.03 & \cellcolor{blue!25}0.25 \\ \hline
Vul & \multicolumn{3}{c||}{R-R-R} & \multicolumn{3}{c||}{R-R-W} & \multicolumn{3}{c||}{R-W-R}& \multicolumn{3}{c||}{R-W-W} & Vul
& \multicolumn{3}{c||}{W-R-R} & \multicolumn{3}{c||}{W-R-W} & \multicolumn{3}{c||}{W-W-R}& \multicolumn{3}{c|}{W-W-W} \\ \hline 
33 & \cellcolor{blue!25}0.55 & \cellcolor{blue!25}0.19 & \cellcolor{blue!25}0.08 & \cellcolor{blue!25}0.41 & \cellcolor{blue!25}0.11 & \cellcolor{blue!25}0.11 & \cellcolor{blue!25}0.41 & \cellcolor{blue!25}0.11 & \cellcolor{blue!25}0.11 & \cellcolor{blue!25}0.42 & \cellcolor{blue!25}0.16 & \cellcolor{blue!25}0.15 & 33 & \cellcolor{blue!25}0.05 & \cellcolor{blue!25}0.02 & \cellcolor{blue!25}0.25 & \cellcolor{blue!25}0.15 & \cellcolor{blue!25}0.02 & 0.00 & \cellcolor{blue!25}0.56 & \cellcolor{blue!25}0.04 & \cellcolor{blue!25}0.01 & \cellcolor{blue!25}0.19 & \cellcolor{blue!25}0.08 & \cellcolor{blue!25}0.34 \\ \hline
34 & \cellcolor{blue!25}0.35 & \cellcolor{blue!25}0.14 & \cellcolor{blue!25}0.10 & \cellcolor{blue!25}0.51 & \cellcolor{blue!25}0.08 & \cellcolor{blue!25}0.02 & \cellcolor{blue!25}0.49 & \cellcolor{blue!25}0.13 & \cellcolor{blue!25}0.04 & \cellcolor{blue!25}0.41 & \cellcolor{blue!25}0.11 & \cellcolor{blue!25}0.11 & 34 & \cellcolor{blue!25}0.18 & \cellcolor{blue!25}0.04 & \cellcolor{blue!25}0.18 & \cellcolor{blue!25}0.10 & \cellcolor{blue!25}0.04 & \cellcolor{blue!25}0.34 & \cellcolor{blue!25}0.09 & \cellcolor{blue!25}0.03 & \cellcolor{blue!25}0.25 & \cellcolor{blue!25}0.34 & \cellcolor{blue!25}0.10 & \cellcolor{blue!25}0.11 \\ \hline
35 & \cellcolor{blue!25}0.24 & 0.00 & 0.00 & \cellcolor{blue!25}0.23 & 0.00 & 0.00 & \cellcolor{blue!25}0.31 & 0.00 & 0.00 & \cellcolor{blue!25}0.32 & 0.00 & 0.00 & 35 & \cellcolor{blue!25}0.12 & 0.00 & 0.00 & \cellcolor{blue!25}0.30 & 0.00 & 0.00 & \cellcolor{blue!25}0.29 & 0.00 & 0.00 & \cellcolor{blue!25}0.39 & 0.00 & 0.00 \\ \hline
36 & 0.00 & \cellcolor{blue!25}0.01 & 0.00 & 0.00 & 0.00 & 0.00 & 0.00 & 0.00 & 0.00 & 0.00 & \cellcolor{blue!25}0.01 & 0.00 & 36 & 0.00 & 0.00 & 0.00 & 0.00 & 0.00 & 0.00 & 0.00 & \cellcolor{blue!25}0.01 & 0.00 & 0.00 & \cellcolor{blue!25}0.01 & 0.00 \\ \hline
37 & \cellcolor{blue!25}0.24 & 0.00 & 0.00 & \cellcolor{blue!25}0.23 & 0.00 & 0.00 & \cellcolor{blue!25}0.31 & 0.00 & 0.00 & \cellcolor{blue!25}0.32 & 0.00 & 0.00 & 37 & \cellcolor{blue!25}0.12 & 0.00 & 0.00 & \cellcolor{blue!25}0.30 & 0.00 & 0.00 & \cellcolor{blue!25}0.29 & 0.00 & 0.00 & \cellcolor{blue!25}0.39 & 0.00 & 0.00 \\ \hline
38 & 0.00 & \cellcolor{blue!25}0.01 & 0.00 & 0.00 & 0.00 & 0.00 & 0.00 & 0.00 & 0.00 & 0.00 & \cellcolor{blue!25}0.01 & 0.00 & 38 & 0.00 & 0.00 & 0.00 & 0.00 & 0.00 & 0.00 & 0.00 & \cellcolor{blue!25}0.01 & 0.00 & 0.00 & \cellcolor{blue!25}0.01 & 0.00 \\ \hline
39 & \cellcolor{blue!25}0.24 & 0.00 & 0.00 & \cellcolor{blue!25}0.23 & 0.00 & 0.00 & \cellcolor{blue!25}0.31 & 0.00 & 0.00 & \cellcolor{blue!25}0.32 & 0.00 & 0.00 & 39 & \cellcolor{blue!25}0.12 & 0.00 & 0.00 & \cellcolor{blue!25}0.30 & 0.00 & 0.00 & \cellcolor{blue!25}0.29 & 0.00 & 0.00 & \cellcolor{blue!25}0.39 & 0.00 & 0.00 \\ \hline
40 & 0.00 & \cellcolor{blue!25}0.01 & 0.00 & 0.00 & 0.00 & 0.00 & 0.00 & 0.00 & 0.00 & 0.00 & \cellcolor{blue!25}0.01 & 0.00 & 40 & 0.00 & 0.00 & 0.00 & 0.00 & 0.00 & 0.00 & 0.00 & \cellcolor{blue!25}0.01 & 0.00 & 0.00 & \cellcolor{blue!25}0.01 & 0.00 \\ \hline
41 & \cellcolor{blue!25}0.09 & \cellcolor{blue!25}0.03 & \cellcolor{blue!25}0.30 & \cellcolor{blue!25}0.13 & \cellcolor{blue!25}0.04 & \cellcolor{blue!25}0.32 & \cellcolor{blue!25}0.10 & \cellcolor{blue!25}0.04 & \cellcolor{blue!25}0.34 & \cellcolor{blue!25}0.09 & \cellcolor{blue!25}0.03 & \cellcolor{blue!25}0.25 & 41 & \cellcolor{blue!25}0.10 & \cellcolor{blue!25}0.04 & \cellcolor{blue!25}0.34 & \cellcolor{blue!25}0.07 & \cellcolor{blue!25}0.03 & \cellcolor{blue!25}0.52 & \cellcolor{blue!25}0.10 & \cellcolor{blue!25}0.04 & \cellcolor{blue!25}0.34 & \cellcolor{blue!25}0.09 & \cellcolor{blue!25}0.03 & \cellcolor{blue!25}0.25 \\ \hline
42 & \cellcolor{blue!25}0.10 & \cellcolor{blue!25}0.04 & \cellcolor{blue!25}0.34 & \cellcolor{blue!25}0.07 & \cellcolor{blue!25}0.03 & \cellcolor{blue!25}0.52 & \cellcolor{blue!25}0.15 & \cellcolor{blue!25}0.02 & 0.00 & \cellcolor{blue!25}0.05 & \cellcolor{blue!25}0.02 & \cellcolor{blue!25}0.39 & 42 & \cellcolor{blue!25}0.15 & \cellcolor{blue!25}0.02 & 0.00 & \cellcolor{blue!25}0.15 & \cellcolor{blue!25}0.06 & \cellcolor{blue!25}0.27 & \cellcolor{blue!25}0.20 & \cellcolor{blue!25}0.08 & \cellcolor{blue!25}0.34 & \cellcolor{blue!25}0.50 & \cellcolor{blue!25}0.12 & \cellcolor{blue!25}0.02 \\ \hline
43 & \cellcolor{blue!25}0.24 & 0.00 & 0.00 & \cellcolor{blue!25}0.23 & 0.00 & 0.00 & \cellcolor{blue!25}0.31 & 0.00 & 0.00 & \cellcolor{blue!25}0.32 & 0.00 & 0.00 & 43 & \cellcolor{blue!25}0.12 & 0.00 & 0.00 & \cellcolor{blue!25}0.30 & 0.00 & 0.00 & \cellcolor{blue!25}0.29 & 0.00 & 0.00 & \cellcolor{blue!25}0.39 & 0.00 & 0.00 \\ \hline
44 & 0.00 & \cellcolor{blue!25}0.01 & 0.00 & 0.00 & 0.00 & 0.00 & 0.00 & 0.00 & 0.00 & 0.00 & \cellcolor{blue!25}0.01 & 0.00 & 44 & 0.00 & 0.00 & 0.00 & 0.00 & 0.00 & 0.00 & 0.00 & \cellcolor{blue!25}0.01 & 0.00 & 0.00 & \cellcolor{blue!25}0.01 & 0.00 \\ \hline
\hline

\multicolumn{26}{|c|}{RF (L) Cache} \\ \hline
Vul & \multicolumn{3}{c||}{F-R-R} & \multicolumn{3}{c||}{F-R-W} & \multicolumn{3}{c||}{F-W-R}& \multicolumn{3}{c||}{F-W-W} & Vul
& \multicolumn{3}{c||}{F-R-R} & \multicolumn{3}{c||}{F-R-W} & \multicolumn{3}{c||}{F-W-R}& \multicolumn{3}{c|}{F-W-W} \\ \hline 
1 & \cellcolor{blue!25}0.55 & \cellcolor{blue!25}0.19 & \cellcolor{blue!25}0.08 & \cellcolor{blue!25}0.41 & \cellcolor{blue!25}0.11 & \cellcolor{blue!25}0.11 & \cellcolor{blue!25}0.42 & \cellcolor{blue!25}0.16 & \cellcolor{blue!25}0.15 & \cellcolor{blue!25}0.07 & \cellcolor{blue!25}0.45 & \cellcolor{blue!25}0.16 & 2 & \cellcolor{blue!25}0.55 & \cellcolor{blue!25}0.19 & \cellcolor{blue!25}0.08 & \cellcolor{blue!25}0.41 & \cellcolor{blue!25}0.11 & \cellcolor{blue!25}0.11 & \cellcolor{blue!25}0.42 & \cellcolor{blue!25}0.16 & \cellcolor{blue!25}0.15 & \cellcolor{blue!25}0.54 & \cellcolor{blue!25}0.60 & \cellcolor{blue!25}0.88 \\ \hline
3 & \cellcolor{blue!25}0.48 & \cellcolor{blue!25}0.10 & \cellcolor{blue!25}0.03 & \cellcolor{blue!25}0.46 & \cellcolor{blue!25}0.09 & \cellcolor{blue!25}0.06 & \cellcolor{blue!25}0.34 & \cellcolor{blue!25}0.10 & \cellcolor{blue!25}0.12 & \cellcolor{blue!25}0.13 & \cellcolor{blue!25}0.04 & \cellcolor{blue!25}0.09 & 4 & \cellcolor{blue!25}0.48 & \cellcolor{blue!25}0.10 & \cellcolor{blue!25}0.03 & \cellcolor{blue!25}0.46 & \cellcolor{blue!25}0.09 & \cellcolor{blue!25}0.06 & \cellcolor{blue!25}0.34 & \cellcolor{blue!25}0.10 & \cellcolor{blue!25}0.12 & \cellcolor{blue!25}0.50 & \cellcolor{blue!25}0.12 & \cellcolor{blue!25}0.02 \\ \hline
5 & \cellcolor{blue!25}0.48 & \cellcolor{blue!25}0.10 & \cellcolor{blue!25}0.03 & \cellcolor{blue!25}0.46 & \cellcolor{blue!25}0.09 & \cellcolor{blue!25}0.06 & \cellcolor{blue!25}0.34 & \cellcolor{blue!25}0.10 & \cellcolor{blue!25}0.12 & \cellcolor{blue!25}0.13 & \cellcolor{blue!25}0.04 & \cellcolor{blue!25}0.09 & 6 & \cellcolor{blue!25}0.48 & \cellcolor{blue!25}0.10 & \cellcolor{blue!25}0.03 & \cellcolor{blue!25}0.46 & \cellcolor{blue!25}0.09 & \cellcolor{blue!25}0.06 & \cellcolor{blue!25}0.34 & \cellcolor{blue!25}0.10 & \cellcolor{blue!25}0.12 & \cellcolor{blue!25}0.50 & \cellcolor{blue!25}0.12 & \cellcolor{blue!25}0.02 \\ \hline
7 & \cellcolor{blue!25}0.19 & \cellcolor{blue!25}0.08 & \cellcolor{blue!25}0.29 & \cellcolor{blue!25}0.49 & \cellcolor{blue!25}0.13 & \cellcolor{blue!25}0.04 & \cellcolor{blue!25}0.29 & \cellcolor{blue!25}0.09 & \cellcolor{blue!25}0.05 & \cellcolor{blue!25}0.05 & \cellcolor{blue!25}0.02 & \cellcolor{blue!25}0.39 & 8 & \cellcolor{blue!25}0.19 & \cellcolor{blue!25}0.08 & \cellcolor{blue!25}0.29 & \cellcolor{blue!25}0.49 & \cellcolor{blue!25}0.13 & \cellcolor{blue!25}0.04 & \cellcolor{blue!25}0.29 & \cellcolor{blue!25}0.09 & \cellcolor{blue!25}0.05 & \cellcolor{blue!25}0.61 & \cellcolor{blue!25}0.79 & \cellcolor{blue!25}0.55 \\ \hline
9 & \cellcolor{blue!25}0.19 & \cellcolor{blue!25}0.08 & \cellcolor{blue!25}0.29 & \cellcolor{blue!25}0.49 & \cellcolor{blue!25}0.13 & \cellcolor{blue!25}0.04 & \cellcolor{blue!25}0.29 & \cellcolor{blue!25}0.09 & \cellcolor{blue!25}0.05 & \cellcolor{blue!25}0.05 & \cellcolor{blue!25}0.02 & \cellcolor{blue!25}0.39 & 10 & \cellcolor{blue!25}0.06 & \cellcolor{blue!25}0.07 & 0.00 & \cellcolor{blue!25}0.02 & \cellcolor{blue!25}0.11 & 0.00 & \cellcolor{blue!25}0.33 & \cellcolor{blue!25}0.71 & \cellcolor{blue!25}0.09 & \cellcolor{blue!25}0.51 & \cellcolor{blue!25}0.92 & \cellcolor{blue!25}0.34\\ \hline
Vul & \multicolumn{3}{c||}{R-R-R} & \multicolumn{3}{c||}{R-R-W} & \multicolumn{3}{c||}{R-W-R}& \multicolumn{3}{c||}{R-W-W} & Vul
& \multicolumn{3}{c||}{W-R-R} & \multicolumn{3}{c||}{W-R-W} & \multicolumn{3}{c||}{W-W-R}& \multicolumn{3}{c|}{W-W-W} \\ \hline 
33 & \cellcolor{blue!25}0.06 & \cellcolor{blue!25}0.07 & 0.00 & \cellcolor{blue!25}0.33 & \cellcolor{blue!25}0.71 & \cellcolor{blue!25}0.09 & \cellcolor{blue!25}0.02 & \cellcolor{blue!25}0.11 & 0.00 & \cellcolor{blue!25}0.04 & \cellcolor{blue!25}0.12 & 0.00 & 33 & \cellcolor{blue!25}0.51 & \cellcolor{blue!25}0.92 & \cellcolor{blue!25}0.34 & \cellcolor{blue!25}0.09 & \cellcolor{blue!25}0.03 & \cellcolor{blue!25}0.30 & \cellcolor{blue!25}0.50 & \cellcolor{blue!25}0.06 & \cellcolor{blue!25}0.03 & \cellcolor{blue!25}0.84 & \cellcolor{blue!25}0.21 & \cellcolor{blue!25}0.05 \\ \hline
34 & \cellcolor{blue!25}0.08 & \cellcolor{blue!25}0.16 & 0.00 & \cellcolor{blue!25}0.16 & \cellcolor{blue!25}0.18 & \cellcolor{blue!25}0.01 & \cellcolor{blue!25}0.26 & \cellcolor{blue!25}0.07 & 0.00 & \cellcolor{blue!25}0.05 & \cellcolor{blue!25}0.38 & \cellcolor{blue!25}0.01 & 34 & \cellcolor{blue!25}0.19 & \cellcolor{blue!25}0.07 & 0.00 & \cellcolor{blue!25}0.11 & \cellcolor{blue!25}0.24 & \cellcolor{blue!25}0.01 & \cellcolor{blue!25}0.34 & \cellcolor{blue!25}0.06 & \cellcolor{blue!25}0.02 & \cellcolor{blue!25}0.67 & \cellcolor{blue!25}0.11 & \cellcolor{blue!25}0.01 \\ \hline
35 & \cellcolor{blue!25}0.14 & \cellcolor{blue!25}0.01 & \cellcolor{blue!25}0.03 & \cellcolor{blue!25}0.06 & \cellcolor{blue!25}0.02 & \cellcolor{blue!25}0.21 & \cellcolor{blue!25}0.08 & \cellcolor{blue!25}0.02 & \cellcolor{blue!25}0.13 & \cellcolor{blue!25}0.09 & \cellcolor{blue!25}0.02 & \cellcolor{blue!25}0.05 & 35 & \cellcolor{blue!25}0.10 & \cellcolor{blue!25}0.04 & \cellcolor{blue!25}0.36 & \cellcolor{blue!25}0.09 & \cellcolor{blue!25}0.02 & \cellcolor{blue!25}0.14 & \cellcolor{blue!25}0.05 & \cellcolor{blue!25}0.02 & \cellcolor{blue!25}0.21 & \cellcolor{blue!25}0.50 & \cellcolor{blue!25}0.12 & \cellcolor{blue!25}0.02 \\ \hline
36 & \cellcolor{blue!25}0.35 & \cellcolor{blue!25}0.14 & \cellcolor{blue!25}0.10 & \cellcolor{blue!25}0.51 & \cellcolor{blue!25}0.08 & \cellcolor{blue!25}0.02 & \cellcolor{blue!25}0.49 & \cellcolor{blue!25}0.13 & \cellcolor{blue!25}0.04 & \cellcolor{blue!25}0.29 & \cellcolor{blue!25}0.09 & \cellcolor{blue!25}0.05 & 36 & \cellcolor{blue!25}0.34 & \cellcolor{blue!25}0.06 & \cellcolor{blue!25}0.02 & \cellcolor{blue!25}0.67 & \cellcolor{blue!25}0.11 & \cellcolor{blue!25}0.01 & \cellcolor{blue!25}0.50 & \cellcolor{blue!25}0.12 & \cellcolor{blue!25}0.02 & \cellcolor{blue!25}0.34 & \cellcolor{blue!25}0.06 & \cellcolor{blue!25}0.02 \\ \hline
37 & \cellcolor{blue!25}0.14 & \cellcolor{blue!25}0.01 & \cellcolor{blue!25}0.03 & \cellcolor{blue!25}0.06 & \cellcolor{blue!25}0.02 & \cellcolor{blue!25}0.21 & \cellcolor{blue!25}0.08 & \cellcolor{blue!25}0.02 & \cellcolor{blue!25}0.13 & \cellcolor{blue!25}0.09 & \cellcolor{blue!25}0.02 & \cellcolor{blue!25}0.05 & 37 & \cellcolor{blue!25}0.10 & \cellcolor{blue!25}0.04 & \cellcolor{blue!25}0.36 & \cellcolor{blue!25}0.09 & \cellcolor{blue!25}0.02 & \cellcolor{blue!25}0.14 & \cellcolor{blue!25}0.05 & \cellcolor{blue!25}0.02 & \cellcolor{blue!25}0.21 & \cellcolor{blue!25}0.67 & \cellcolor{blue!25}0.11 & \cellcolor{blue!25}0.01 \\ \hline
38 & \cellcolor{blue!25}0.35 & \cellcolor{blue!25}0.14 & \cellcolor{blue!25}0.10 & \cellcolor{blue!25}0.51 & \cellcolor{blue!25}0.08 & \cellcolor{blue!25}0.02 & \cellcolor{blue!25}0.49 & \cellcolor{blue!25}0.13 & \cellcolor{blue!25}0.04 & \cellcolor{blue!25}0.29 & \cellcolor{blue!25}0.09 & \cellcolor{blue!25}0.05 & 38 & \cellcolor{blue!25}0.34 & \cellcolor{blue!25}0.06 & \cellcolor{blue!25}0.02 & \cellcolor{blue!25}0.67 & \cellcolor{blue!25}0.11 & \cellcolor{blue!25}0.01 & \cellcolor{blue!25}0.50 & \cellcolor{blue!25}0.12 & \cellcolor{blue!25}0.02 & \cellcolor{blue!25}0.59 & \cellcolor{blue!25}0.17 & \cellcolor{blue!25}0.08 \\ \hline
39 & \cellcolor{blue!25}0.14 & \cellcolor{blue!25}0.01 & \cellcolor{blue!25}0.03 & \cellcolor{blue!25}0.06 & \cellcolor{blue!25}0.02 & \cellcolor{blue!25}0.21 & \cellcolor{blue!25}0.08 & \cellcolor{blue!25}0.02 & \cellcolor{blue!25}0.13 & \cellcolor{blue!25}0.09 & \cellcolor{blue!25}0.02 & \cellcolor{blue!25}0.05 & 39 & \cellcolor{blue!25}0.10 & \cellcolor{blue!25}0.04 & \cellcolor{blue!25}0.36 & \cellcolor{blue!25}0.09 & \cellcolor{blue!25}0.02 & \cellcolor{blue!25}0.14 & \cellcolor{blue!25}0.05 & \cellcolor{blue!25}0.02 & \cellcolor{blue!25}0.21 & \cellcolor{blue!25}0.84 & \cellcolor{blue!25}0.21 & \cellcolor{blue!25}0.05 \\ \hline
40 & \cellcolor{blue!25}0.13 & \cellcolor{blue!25}0.04 & \cellcolor{blue!25}0.32 & \cellcolor{blue!25}0.14 & \cellcolor{blue!25}0.05 & \cellcolor{blue!25}0.27 & \cellcolor{blue!25}0.59 & \cellcolor{blue!25}0.17 & \cellcolor{blue!25}0.08 & \cellcolor{blue!25}0.09 & \cellcolor{blue!25}0.03 & \cellcolor{blue!25}0.30 & 40 & \cellcolor{blue!25}0.50 & \cellcolor{blue!25}0.06 & \cellcolor{blue!25}0.03 & \cellcolor{blue!25}0.84 & \cellcolor{blue!25}0.21 & \cellcolor{blue!25}0.05 & \cellcolor{blue!25}0.13 & \cellcolor{blue!25}0.04 & \cellcolor{blue!25}0.32 & \cellcolor{blue!25}0.14 & \cellcolor{blue!25}0.05 & \cellcolor{blue!25}0.27 \\ \hline
41 & \cellcolor{blue!25}0.59 & \cellcolor{blue!25}0.17 & \cellcolor{blue!25}0.08 & \cellcolor{blue!25}0.10 & \cellcolor{blue!25}0.04 & \cellcolor{blue!25}0.34 & \cellcolor{blue!25}0.11 & \cellcolor{blue!25}0.05 & \cellcolor{blue!25}0.38 & \cellcolor{blue!25}0.06 & \cellcolor{blue!25}0.03 & \cellcolor{blue!25}0.49 & 41 & \cellcolor{blue!25}0.07 & \cellcolor{blue!25}0.03 & \cellcolor{blue!25}0.52 & \cellcolor{blue!25}0.04 & \cellcolor{blue!25}0.01 & \cellcolor{blue!25}0.18 & \cellcolor{blue!25}0.09 & \cellcolor{blue!25}0.03 & \cellcolor{blue!25}0.35 & \cellcolor{blue!25}0.09 & \cellcolor{blue!25}0.03 & \cellcolor{blue!25}0.25 \\ \hline
42 & \cellcolor{blue!25}0.10 & \cellcolor{blue!25}0.04 & \cellcolor{blue!25}0.34 & \cellcolor{blue!25}0.11 & \cellcolor{blue!25}0.05 & \cellcolor{blue!25}0.38 & \cellcolor{blue!25}0.06 & \cellcolor{blue!25}0.03 & \cellcolor{blue!25}0.49 & \cellcolor{blue!25}0.07 & \cellcolor{blue!25}0.03 & \cellcolor{blue!25}0.52 & 42 & \cellcolor{blue!25}0.04 & \cellcolor{blue!25}0.01 & \cellcolor{blue!25}0.18 & \cellcolor{blue!25}0.09 & \cellcolor{blue!25}0.03 & \cellcolor{blue!25}0.35 & \cellcolor{blue!25}0.09 & \cellcolor{blue!25}0.03 & \cellcolor{blue!25}0.25 & \cellcolor{blue!25}0.10 & \cellcolor{blue!25}0.04 & \cellcolor{blue!25}0.34 \\ \hline 
43 & \cellcolor{blue!25}0.11 & \cellcolor{blue!25}0.05 & \cellcolor{blue!25}0.38 & \cellcolor{blue!25}0.06 & \cellcolor{blue!25}0.03 & \cellcolor{blue!25}0.49 & \cellcolor{blue!25}0.07 & \cellcolor{blue!25}0.03 & \cellcolor{blue!25}0.52 & \cellcolor{blue!25}0.04 & \cellcolor{blue!25}0.01 & \cellcolor{blue!25}0.18 & 43 & \cellcolor{blue!25}0.09 & \cellcolor{blue!25}0.03 & \cellcolor{blue!25}0.35 & \cellcolor{blue!25}0.09 & \cellcolor{blue!25}0.03 & \cellcolor{blue!25}0.25 & \cellcolor{blue!25}0.55 & \cellcolor{blue!25}0.19 & \cellcolor{blue!25}0.08 & \cellcolor{blue!25}0.34 & \cellcolor{blue!25}0.06 & \cellcolor{blue!25}0.02 \\ \hline

    \end{tabular}
    \end{center}
\end{table*}

Table~\ref{tbl_p_value} lists the {\em pvalue} evaluation results for the normal caches and other secure caches
shown in Figure~\ref{fig:norm_se_rf}.
We assign the blue color to the corresponding entry if the {\em pvalue} result
is larger than $0.00049$, which is the threshold of the {\em pvalue} in our tests to
determine if different timing distributions are distinguishable.
Three  values  of one specific vulnerability for a certain cache  are the {\em pvalue}
for each two timing distributions out of three. 
If it is smaller than $0.00049$, we judge the two timing distributions to be differentiable, otherwise not.
If a vulnerability is effective, at least one timing distributions should be differentiable from the other two.
From the {\em pvalue}  results it can be seen that Table~\ref{tbl_p_value}
and Figure~\ref{fig:norm_se_rf} show the same secure cache evaluation results.
Normal cache cannot prevent all of the vulnerabilities.
PL cache has problems preventing vulnerabilities implemented using store due to the store buffer.
Setting the window size to be small for RF cache will only prevent some vulnerabilities while
a larger window size of RF cache is able to  prevent all the vulnerabilities.

\bibliographystyle{IEEEtran}
\bibliography{refs}

\begin{IEEEbiographynophoto}{Shuwen Deng}
(S'18)
received her B.Sc. in Microelectronics from Shanghai Jiao Tong	University in 2016. She is currently a Ph.D. candidate at the department of Electrical Engineering at Yale University, working with Prof. Jakub Szefer. Her current research includes developing and verifying secure processor microarchitectures by self-developing timing side-channel vulnerability checking schemes, as well as proposing languages and tools for practical and scalable security hardware and architectures verification.
\end{IEEEbiographynophoto}
\vspace{-36pt}
\begin{IEEEbiographynophoto}{Nikolay Matyunin}
(S'20) received his Dipl. in Computer Science from the Lomonosov Moscow State University in 2014. He is currently a Ph.D. candidate at the department of Computer Science at Technical University of Darmstadt, Germany, working with Prof. Dr. Stefan Katzenbeisser. His research interests include covert channels and physical side-channel attacks, security and privacy of mobile and embedded systems.
\end{IEEEbiographynophoto}
\vspace{-36pt}
\begin{IEEEbiographynophoto}{Wenjie Xiong}
(S’17-M’21) received the BSc degree in microelectronics and psychology from Peking University, China, in 2014, 
and the PhD degree from the Department of Electrical Engineering, Yale University, New Haven, Connecticut, in 2020. 
Her research interests comprise physically unclonable functions and side-channel attacks and defenses.
\end{IEEEbiographynophoto}
\vspace{-36pt}
\begin{IEEEbiographynophoto}{Stefan Katzenbeisser}
(S'98--A'01--M'07--SM'12) received the Ph.D. degree from the Vienna University of Technology, Austria. After working as a Research Scientist with the Technical University of Munich, Germany, he joined Philips Research as a Senior Scientist in 2006.
After holding a professorship for Security Engineering at the Technical University of Darmstadt, he joined University of Passau in 2019, heading the Chair of Computer Engineering. His current research interests include embedded security, data privacy and cryptographic protocol design. 
\end{IEEEbiographynophoto}
\vspace{-36pt}
\begin{IEEEbiographynophoto}{Jakub Szefer}
(S'08--M'13--SM'19) received B.S. with highest honors in Electrical and Computer Engineering from University of Illinois at Urbana-Champaign, and  M.A. and Ph.D. degrees in Electrical Engineering from Princeton University where he worked with Prof. Ruby B. Lee on secure hardware architectures.  He is currently an Associate Professor in the Electrical Engineering department at Yale University, where he leads the Computer Architecture and Security Laboratory (CASLAB).  His research interests are at the intersection of computer architecture, hardware security, and FPGA security. 
\end{IEEEbiographynophoto}

\end{document}